\documentclass[sn-mathphys,numbered]{sn-jnl}

\usepackage{mathptmx}

\usepackage{amsmath,amssymb,amsxtra,amsfonts,cancel,mathrsfs}
\usepackage{amsthm}
\usepackage{xspace}
\usepackage{multirow}%
\usepackage{color}
\usepackage{xurl}
\usepackage{caption}
\usepackage{subcaption}
\usepackage{graphicx}
\usepackage{lipsum}
\usepackage[skip=0pt]{caption}
\usepackage{textcomp}%
\usepackage{manyfoot}%
\usepackage{algorithm}%
\usepackage{algorithmicx}%
\usepackage{algpseudocode}%

\usepackage{multirow}
\usepackage{tikz}
\usepackage{fancyvrb}
\usepackage{placeins}
\usepackage{tabularx}

\usepackage{todonotes}
\usepackage{listings}
\DeclareMathOperator*{\argmin}{argmin}
\theoremstyle{thmstyleone}%
\theoremstyle{thmstyletwo}%
\newtheorem{example}{Example}%

\theoremstyle{thmstylethree}%
\newtheorem{definition}{Definition}%

\newcommand{\powerset}[1]{\ensuremath{2^{#1}}} 
\newcommand{\AF}{\ensuremath{\mathcal{F}}\xspace} 
\newcommand{\F}{\ensuremath{\mathcal{F}}\xspace} 
\newcommand{\args}{\ensuremath{\mathsf{A}}\xspace} 
\newcommand{\atts}{\ensuremath{\rightarrow}\xspace}
\newcommand{\AFC}{\ensuremath{\AF=(\args,\atts)}\xspace} 

\newcommand{\af}{AF}

\newcommand{\cf}{\mathbf{cf}}
\newcommand{\ad}{\mathbf{ad}}
\newcommand{\co}{\mathbf{co}}
\newcommand{\pr}{\mathbf{pr}}
\newcommand{\st}{\mathbf{st}}
\newcommand{\sst}{\mathbf{sst}}
\newcommand{\stg}{\mathbf{stg}}
\newcommand{\gr}{\mathbf{gr}}
\newcommand{\id}{\mathbf{id}}
\newcommand{\eg}{\mathbf{eg}}

\newcommand{\se}{\mathbf{SE}}

\newcommand{\dc}{\mathbf{DC}}
\newcommand{\ds}{\mathbf{DS}}

\newcommand{\ex}{\mathbf{EX}}
\newcommand{\nem}{\mathbf{NE}}

\newcommand{\BA}{\textbf{B1}\xspace}
\newcommand{\BB}{\textbf{B2}\xspace}
\newcommand{\BC}{\textbf{B3}\xspace}
\newcommand{\BD}{\textbf{B4}\xspace}

\begin{document}

\title{An Encoding of  Argumentation Problems Using Quadratic Unconstrained Binary Optimization}

\author*[1]{\fnm{Marco} \sur{Baioletti}}\email{marco.baioletti@unipg.it}

\author*[1]{\fnm{Francesco} \sur{Santini}}\email{francesco.santini@unipg.it}

\affil*[1]{\orgdiv{Dipartimento di Matematica e Informatica}, \orgname{Universit{\`a} degli Studi di Perugia}, \orgaddress{\street{Via Vanvitelli 1}, \city{Perugia}, \postcode{06123}, \country{Italy}}}

\abstract{In this paper, we develop a way to encode several NP-Complete problems in \emph{Abstract Argumentation} to \emph{Quadratic Unconstrained Binary Optimization} (\emph{QUBO}) problems. In this form, a solution for a QUBO problem involves minimizing a quadratic function over binary variables ($0$/$1$), where the coefficients can be represented by a symmetric square matrix (or an equivalent upper triangular version). With the QUBO formulation, exploiting new computing architectures, such as Quantum and Digital Annealers, is possible. A more conventional approach consists of developing approximate solvers, which, in this case, are used to tackle the intrinsic complexity. We performed tests to prove the correctness and applicability of classical problems in Argumentation and enforcement of argument sets. We compared our approach to two other approximate solvers in the literature during tests. In the final experimentation, we used a Simulated Annealing algorithm on a local machine. Also, we tested a Quantum Annealer from the D-Wave Ocean SDK and the Leap\textsuperscript{\texttrademark} Quantum Cloud Service.}

\keywords{Argumentation, Abstract Frameworks, Quadratic Unconstrained Binary Optimization, Approximate solvers, Quantum Annealers}



\maketitle

\section{Introduction}\label{sec:introduction}
\emph{Computational Argumentation}\footnote{Whose community has as one of its main references in the ``International Conference on Computational Models of Argument'' conference series, since the year 2006.} is an interdisciplinary field that combines concepts from Artificial Intelligence, Logic, Philosophy, and Cognitive Science to study and develop systems that can model, analyze, and generate arguments~\cite{baroni2018handbook,gabbay2021handbook}.

A good share of the research on the subject of \emph{Argumentation}~\cite{pollock,simari,vreeswijk} is based on the theory initiated by Dung in 1995~\cite{Dung:1995}. There, an \emph{Abstract Argumentation Framework} (hereafter \emph{AF} for short) can be visually represented as a directed graph, formally a tuple $\F=(\args,\atts)$: nodes point to arguments (the set $\args$), and edges (the set $\atts$) represent a directed ``attack'' relation between two arguments (i.e., $\atts: \args \times \args$), in which an argument rebuts another argument. Given such a network, an analysis of which set(s) of arguments can be collectively accepted leads to the definition of \emph{semantics}. Semantics declaratively rule which coherent sets of arguments can be acceptable for an agent participating in a debate; these sets of arguments are named \emph{extensions}. Extensions are based on two fundamental concepts: conflict-freeness (an extension cannot contain two arguments in conflict), and \emph{admissibility}: a set of arguments is admissible if any of its arguments counter-attacks each attacker of its elements, that is, a set ``defends'' all its arguments. For example, a \emph{preferred} extension is the maximal element (concerning set-theoretical inclusion) among all the possible admissible sets in an AF.  Strengthening the basic requirements enforced by admissibility yields complete semantics~\cite{Baroni:2011}: admissibility refers to the ability to provide reasons for accepting or rejecting arguments but allows individuals to abstain from deciding on any argument. Complete semantics requires individuals to abstain only when there are no convincing reasons to do otherwise. Maximizing the number of arguments accepted is a concept captured by \emph{preferred} semantics, while in the \emph{stable} semantics, every argument that is not accepted is rejected~\cite{Baroni:2011}.

In the AF-based model proposed by \cite{Dung:1995}, arguments have no internal structure. Consequently, they are not composed of premises and claims, as it happens instead when Argumentation is studied at lower levels of abstraction. Moreover, the attack relationship also represents a generic notion of conflict and is not connected to any form of logical negation. As an example of debate, two conflicting arguments could be \emph{a: ``I should buy an electric car because it is environmentally friendly''} and \emph{b: ``Electric cars are not so environmentally friendly if the power used to recharge them comes from fossils''}, and in this case, $b \atts a$.

The paper's second major characteristic involves \emph{Quadratic Unconstrained Binary Optimization} problems (\emph{QUBO}),\footnote{Different names and abbreviations may be found in the literature, as \emph{Unconstrained Binary Quadratic Programming} (\emph{UBPQ})~\cite{survey1}, or \emph{Quadratic Pseudo-Boolean} optimization (\emph{QPBO})~\cite{survey2}.} which correspond to a mathematical formulation that encompasses a wide range of critical \emph{Combinatorial Optimization} problems: QUBO has been surveyed in \cite{survey2,survey1}, and the first work dates back to 1960~\cite{firstworkqubo}. A solution to a QUBO problem consists of minimizing a quadratic function over binary variables (that is, $0$ /$1$), whose coefficients can be represented with a symmetric square matrix or in an equivalent upper triangular form achieved without loss of generality. QUBO problems are NP-Complete: for this reason, a vast literature is dedicated to approximate solvers based on heuristics or metaheuristics, such as \emph{simulated annealing} approaches (\emph{SA}), \emph{tabu-serch}, \emph{genetic algorithms} or \emph{evolutionary computing}~\cite{survey1}. There are also exact methods capable of solving QUBO problems with $100$-$500$ variables~\cite{survey1}. 

QUBO is also the mathematical model accepted by quantum and digital annealers. Hence, a QUBO formulation of an optimization problem is important because it allows the use of these new alternative computing devices.

In this paper, we propose an encoding to QUBO of some classical problems in Argumentation, such as: 

\begin{itemize}
	\item checking if a given argument $a$ is accepted in at least one complete, preferred, and stable extension (credulous acceptance of $a$);
	\item checking the existence of at least one stable extension;
	\item checking the existence of a non-empty extension when using the complete, preferred, semi-stable, and stable semantics;
        \item minimizing the number of modifications to an AF needed for a given set of arguments $T$ to satisfy a given semantics (called \emph{strict enforcement} of $T$~\cite{extenf}).
\end{itemize}

We focus on these tasks because they are all well-known NP-Complete problems in Argumentation and show the same complexity class as solving QUBO problems. The goal is to propose a different kind of encoding and to test and use solvers that are totally different from the ones used in the literature, usually based on \emph{reduction-based approaches} (e.g., \emph{SAT}, \emph{CSP}, \emph{ASP} encodings),\footnote{Respectively, \emph{Boolean Satisfiability Problems}, \emph{Constraint Satisfaction Problems}, and \emph{Answer-set Programming}.} and \emph{direct approaches} (i.e., ad-hoc algorithms)~\cite{cerutti2017foundations}. In practice, we code the obtained QUBO encodings by adopting the \emph{D-Wave Ocean SDK}\footnote{D-Wave Ocean Software Documentation:\url{https://docs.ocean.dwavesys.com/en/stable/}.} and use a \emph{Simulated Annealing} (\emph{SA}) algorithm~\cite{annealing} in that SDK to compute a solution to these problems. SA is a probabilistic technique that approximates the global optimum of a function in a large search space.  

Hence, the solver we propose is approximate as the other argumentation solvers with which we compare it on some of the problems mentioned above: these solvers are \emph{AFGCN} and \emph{Harper++} (see Section~\ref{sec:related}. The solver we propose is generally slower but more accurate w.r.t. such related work, which often fails to return the correct answer. 

At the same time,  the D-Wave Ocean SDK can be used to employ real \emph{Quantum Annealers} to find the global minima of the proposed QUBO encoding.  A section of this paper (Section~\ref{sec:implementationQ}) is dedicated to testing the performance obtained for some of the presented problems on a quantum annealer offered by D-Wave, on which we had a limited amount of time for our experiments. Even if we reduced the size of the AFs because larger problems solved locally were not solvable (or just embeddable) on the Quantum Annealer, these tests represent a first attempt for this kind of Argumentation-related problems. 

This paper elaborates on the preliminary works in \cite{pricai22} and \cite{aiqqia} by blending and amalgamating them, providing QUBO encodings of more problems (detailed in Section~\ref{sec:encoding}) and testing such encodings,  and finally comparing with a further approximate solver, i.e., AFGCN A (Section~\ref{sec:comparison}). The paper is structured as follows: Section~\ref{sec:background} reports the introductory notions about Argumentation, Quantum Annealing, and QUBO. Section~\ref{sec:encoding} presents a QUBO encoding for all the (classical) NP-Complete problems in  Argumentation viewed as AFs, and an encoding of the complete semantics concerning the enforcement problems. Section~\ref{sec:implementation} presents an empirical validation of the model by performing tests and comparing the results against an exact and two further approximate solvers; moreover, we also summarize the performance we obtained while (strictly) enforcing argument sets as complete extensions. Section~\ref{sec:implementationQ} tests Argumentation-related problems on a real quantum annealer provided by D-Wave, and draws first considerations. Finally,  Section~\ref{sec:related} reports some of the most related work, while Section~\ref{sec:conclusion} closes the paper with final thoughts and future work.

\section{Background}\label{sec:background}

This section presents the fundamental background concept necessary to understand the rest of this work. Section~\ref{sec:bgabstract}  describes Argumentation~\cite{Dung:1995} by first introducing fundamental notions such as AFs, semantics, and related problems to be solved and then presenting an extension of such problems to enforce extensions. Section~\ref{sec:qubo} comments on QUBO problems and their solutions. 

\subsection{Problems in Abstract Argumentation Frameworks}\label{sec:bgabstract}

An \emph{Abstract Argumentation Framework} (\af, for short) \cite{Dung:1995}
is a tuple $\F=(\args,\atts)$ where
\args is a set of arguments and
\atts is a relation $\atts\subseteq \args\times\args$.
For two arguments $a, b\in\args$, the relation $a \atts b$ means that argument $a$ \emph{attacks} argument $b$.
An argument $a \in \args$ is \emph{defended} by $S \subseteq \args$ (in $\F$)
if for each $b \in \args$, such that $b \atts a$,
there is some $c \in S$ such that $c \atts b$.
A set $E \subseteq \args$ is \emph{conflict-free} ($\cf$ in \F) if and only if there is no $a, b\in E$ with $a \atts b$.
$E$ is \emph{admissible} ($\ad$ in \F) if and only if it is conflict-free and if each $a \in E$ is defended by $E$.
Finally, the range of $E$ in $\F$, i.e., $E^{+}_\F$, collects the same $E$ and the set of arguments attacked by $E$ is $$E^{+}_\F=E \cup \{a\in\args \mid \exists b\in E: b \atts a\}. $$

\begin{figure}[b]
	\centering
	\begin{tikzpicture}[scale=1, transform shape]
		\tikzstyle{every node} = [line width=1pt, shape=rectangle, fill=gray!15, minimum width=0.6cm]
		\node (a) at (3, 0) {$a$};
		\node (b) at +(0: 4.5) {$b$};
		\node (c) at +(0: 6) {$c$};
		\node (d) at +(0: 7.5) {$d$};
		\node (e) at +(0: 9) {$e$};
		\draw [line width = 1pt, ->] (a) -- (b) node[pos=.5, fill=white, above] {};
		\draw [line width = 1pt, ->] (c) -- (b) node[pos=.5, fill=white, above] {};
		\draw [line width = 1pt, ->] (d) -- (e) node[pos=.5, fill=white, above] {};
		\draw [line width = 1pt, ->] (c)  edge[bend right=25]   node[fill=white, below] {} (d);
		\draw [line width = 1pt, ->] (d) edge[bend right=25] node[fill=white, above] {} (c);
	\end{tikzpicture}
	\caption{An example of an AF represented as a directed graph.}\label{fig:argnetex}
\end{figure}
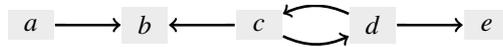

A directed graph can straightforwardly represent an AF: an example with five arguments is given in Figure~\ref{fig:argnetex}: $\F=(\{a,b,c,d,e\},\{a \atts b, c\atts b, c \atts d, d \atts c, d \atts e\})$.

The \emph{collective acceptability} of arguments depends on the definition of different \textit{semantics}. 
Semantics determine sets of jointly acceptable arguments, i.e., sets of arguments called \emph{extensions}, by mapping each \AFC to a set $\sigma(\F) \subseteq \powerset{\args}$, where $\powerset{\args}$ is the power-set of $\args$, and $\sigma$ parametrically stands for any of the considered semantics.

Four semantics were proposed by Dung in his seminal paper~\cite{Dung:1995}, namely the complete ($\co$), preferred ($\pr$), stable ($\st$), and grounded ($\gr$) semantics. Succesive works defined the semi-stable ($\sst$)~\cite{CaminadaCD12}, the stage ($\stg$)~\cite{Verheij:1996}, the ideal ($\id$)~\cite{DungMT:2007}, and finally the eager ($\eg$)~\cite{eager} semantics.
Given \AFC and a set $E \subseteq \args$, we report the definition of all these semantics:

\begin{itemize}
	\item $E \in \co(\F)$ iff $E$ is admissible in $\F$ and if $a \in \args$ is defended by $E$ in $\F$ then $a\in E$,	
	\item $E \in \pr(\F)$ iff $E \in \co(\F)$ and there is no $E' \in \co(\F)$ s.t.\ $E' \supset E$,
	\item $E \in \sst(\F)$ iff $E \in \co(\F)$ and there is no $E' \in \co(\F)$ s.t.\ $E'^+_\F \supset E^+_\F$,
	\item $E \in \st(\F)$ iff $E \in \co(\F)$ and $E^+_\F = \args$,
	\item $E \in \stg(\F)$ iff $E$ is conflict-free in $\F$ and there is no 
	$E'$ that is 
	conflict-free 
	in $\F$ s.t.\ $E'^+_\F \supset E^+_\F$,
	\item $E \in \gr(\F)$ iff $E \in \co(\F)$ and there is no $E' \in \co(\F)$ s.t.\ $E' \subset E$,
	\item $E \in \id(\F)$ if and only if $E$  is admissible, $E \subseteq \bigcap{\pr(\F)}$ and
	there is no admissible $E' \subseteq \bigcap{\pr(\F)}$ s.t.\ $E' \supset E$;
 	\item $E \in \eg(\F)$ if and only if $E$  is admissible, $E \subseteq \bigcap{\sst(\F)}$ and
	there is no admissible $E' \subseteq \bigcap{\sst(\F)}$ s.t.\ $E' \supset E$.
\end{itemize}

For a more detailed view of these semantics, please refer to \cite{Baroni:2011}. Note that the grounded, the ideal, and the eager extensions are uniquely determined~\cite{Dung:1995,DungMT:2007,eager}. Thus, they are also called \emph{single-status} semantics. The other semantics introduced are  \emph{multi-status} semantics, where several extensions may exist. The stable semantics is the only case where $\st(\F)$ might be empty, while at least one extension always satisfies the other semantics.

As an example, if we consider the framework $\F$ in Figure~\ref{fig:argnetex} we have for example that 
\begin{itemize}
\item $\cf(\F) = \{\emptyset, \{a\}, \{b\}, \{c\}, \{d\}, \{e\}, \{a, c\}, \{a, d\}, \{a, e\}, \{b, d\}, \{b, e\}, \{c, e\}, \{a, c, e\}\}$;
\item $\ad(\F) = \{\emptyset, \{a\}, \{c\}, \{d\}, \{a, c\}, \{a, d\}, \{c, e\}, \{a, c, e\}\}$;
\item $\co(\F) = \{\{a\}, \{a, d\}, \{a, c, e\}\}$;
\item $\pr(\F) = \{\{a, d\}, \{a, c, e\}\}$;
\item $ \sst(\F) = \st(\F) = \stg(\F)=\{\{a,d\}, \{a, c, e\}\}$;
\item $\gr(\F) = \id(\F) =  \eg(\F) = \{\{a\}\}$.
\end{itemize}

We now report below the definition of six well-known problems in Argumentation, all of which are decision ones (yes/no answer): 

\begin{itemize}
	\item Credulous acceptance $\dc\textit{-}\sigma$: given \AFC and an argument $a \in \args$, is $a$ contained in some $E \in \sigma(\F)$?
	\item Skeptical acceptance $\ds\textit{-}\sigma$: given \AFC and an argument $a \in \args$, is $a$ contained in all $E \in \sigma(\F)$?
	\item Verification of an extension $\mathit{\textbf{VER}}\textit{-}\sigma$: given \AFC and a set of arguments $E \subseteq \args$, is $E \in \sigma(\F)$?
	\item Existence of an extension $\ex\textit{-}\sigma$: given \AFC, is
	$\sigma(\F) \not= \varnothing$?
	\item Existence of non-empty extension
	$\nem\textit{-}\sigma$: given \AFC, does there exist $E
	\not= \varnothing$ such that $E \in \sigma(\F)$?
	\item Uniqueness of the solution $\mathit{\textbf{UN}}\textit{-}\sigma$: given \AFC, is $\sigma(\F) = \{E\}$?
\end{itemize}

For example, $\dc$-$\co$ for the AF in Figure~\ref{fig:argnetex} returns ``yes'' for argument $c$ and ``no'' for argument $b$; $\ds$-$\co$  returns ``yes'' for argument $a$ only; $\nem\textit{-}\st$ returns ``yes''.

Table~\ref{sec:complexity} summarizes the complexity classes of the problems mentioned above~\cite{dvorakcomplexity}. As we can see, most of
the problems need efficient solvers. As a reminder, intractable complexity
classes are $\mathit{NP}, \mathit{coNP}, \mathit{DP} \subseteq \Theta^{p}_{2}
\subseteq \sum^{P}_{2}, \prod^{P}_{2} \subseteq D^{p}_{2}$, while all the other
classes in Table~\ref{sec:complexity} are tractable: $\mathit{L} \subseteq \mathit{P}$.

In the remaining sections of this paper, our focus will solely be on NP-Complete problems in Table~\ref{sec:complexity} since they have the same complexity as solving QUBO. However, in this section, we have provided a broader perspective for the sake of completeness. We have noted that the stage, ideal, and grounded semantics do not exhibit any NP-Complete problem. Furthermore, the verification problem of any presented semantics does not imply NP-Completeness.

Since Dung's seminal paper in 1995, \cite{Dung:1995}, AFs have been expanded in various ways, such as by adding weights to the attack relation.~\cite{dunne,japprox18}, probability values~\cite{epistemic,oren}, adding claims to arguments~\cite{claim}, making it possible to have attacks on attacks~\cite{attacksbaroni}, or showing incompleteness in the structure of the frameworks~\cite{incomplete1,incomplete2}.\footnote{This paper only presents a subset of possible proposed extensions of \cite{Dung:1995} A complete survey paper would be outside the scope.} In the next section, we will discuss an approach closely related to the issues we presented earlier. This approach effectively captures the dynamic nature that characterizes debates.

\begin{table*}[]
\centering
\small
\begin{tabular}{ccccccc}
&  Ver$_\sigma$ &  Cred$_\sigma$ & Scept$_\sigma$ & Exists$_\sigma$ & Exists$_\sigma^{\neg\varnothing}$ & Unique$_\sigma$ \\
Conflict-free & in L & in L & triv.& triv. & in L & in L \\
Admissible & in L  & \bf{NP-c} & triv. & triv. &  \bf{NP-c} & coNP-c \\
Complete & in L &  \bf{NP-c} & P-c & triv. &  \bf{NP-c} & coNP-c \\
Preferred & coNP-c &  \bf{NP-c} & $\prod^{P}_{2}$-c& triv. &  \bf{NP-c} & coNP-c\\
Semi-stable & coNP-c & $\sum^{P}_{2}$-c & $\prod^{P}_{2}$-c & triv. &  \bf{NP-c} & in $\Theta^{p}_{2}$ \\
Stable & in L &  \bf{NP-c} & coNP-c &  \bf{NP-c} &  \bf{NP-c} & DP-cP\\
Stage & coNP-c & $\sum^{P}_{2}$-c & $\prod^{P}_{2}$-c & triv. & in L & in $\Theta^{p}_{2}$ \\
Grounded & P-c & P-c & P-c & triv. & P-c & triv. \\
Ideal & $\Theta^{p}_{2}$ & $\Theta^{p}_{2}$ & $\Theta^{p}_{2}$ & triv. & $\Theta^{p}_{2}$ & triv. \\
Eager & $D^{p}_{2}$ & $\prod^{P}_{2}$-c & $\prod^{P}_{2}$-c & triv. & $\prod^{P}_{2}$-c & triv.
\end{tabular}
\caption{The complexity of some classical problems in Argumentation~\cite{dvorakcomplexity}. In bold, we highlight the problems with the same complexity as solving QUBO, i.e., NP-Complete problems.}
\label{sec:complexity}
\end{table*}

\subsubsection{Extension Enforcement Problems}\label{sect:bgenforcement}
In a multi-agent system, AFs are used to represent and reason about different schemes of dialogues (e.g., persuasion- or negotiation-oriented dialogues). Regardless of the subject, it is crucial to understand the dynamics of debates and, consequently, AFs and how to update them in response to new information. This problem has been studied for the first time in \cite{baumann1,baumann2}: these works enforce a set of arguments by adding new arguments and some attacks between such new arguments; attacks need to satisfy some constraints, for example, in a \emph{strong} expansion~\cite{baumann1}, no new attack is directed from a former argument to a new one. However, no change to $\atts$ only is allowed in such papers.

Along the same line of research, the work in \cite{extenf1} presents the task of \emph{extension enforcement}: the authors consider the objective to change the attack relationship $\atts$ of a framework $\F=(\args,\atts)$ such that a given set $T \subseteq \args$ becomes (a subset of) an extension under a given semantics $\sigma$. In this case, the enforcement is \emph{argument-fixed} since only the attack relationship can be modified by adding or removing arguments. This form of enforcement is generally characterized by two distinct levels of requirements: \emph{Strict enforcement} is satisfied if $T$ is precisely a $\sigma$-extension. In contrast, in \emph{non-strict enforcement} $T$ is only required to be a subset of a $\sigma$-extension.  Enforcement operators are helpful in many scenarios, particularly when multiple agents are involved in heated debates. In such debates, new arguments often challenge previously accepted extensions, and an agent may feel compelled to develop new arguments to support their preferred view. However, in numerous other situations, no new argument can be used to explain the change, and thus, differently from \cite{baumann1,baumann2}, only the attack relationship can be modified.

When considering the Hamming distance of the changes, i.e., $|\atts \Delta \atts'| = |\atts\setminus \atts'|+|\atts'\setminus \atts|$, in \cite{extenf1} the authors impose a threshold $|\atts \Delta \atts'| \leq k$ as a further parameter of these problems. The complexity of some of these problems is reported in Table~\ref{tab:complexity2}: Even in this case, some problems are characterized by NP-Completeness and are consequently predisposed to be solved in a QUBO format. 

In this paper, as proposed in \cite{extenf}, we look at the problem from an optimization point of view by minimizing the number of modifications (i.e., addition and removal of attacks) needed on an AF to make a set of arguments (or a subset) satisfy a given semantics on the modified AF:

\begin{definition}[Extension enforcement~\cite{extenf}]
	Given a framework $\F=(\args,\atts)$, a set of arguments $T \subseteq \args$, and a semantics $\sigma$, strict extension enforcement is an optimization problem where the goal is to find $\F^\ast=(\args,\atts^\ast)$ s.t.:
	$$\atts^\ast \in \argmin_{\atts' \in \mathit{enf_{st}}(\F, T, \sigma)} \; |\atts \Delta \atts'|$$
	\noindent where $\mathit{enf_{st}}(\F, T, \sigma)= \{\atts' | \F' =(\args,\atts'), T \in \sigma(\F')\}$ is the strict enforcement relation: $T \in \sigma(\F^\ast)$. Similarly, we can define the same problem by considering non-strict enforcements (by defining $\mathit{enf_{nst}}(\F, T, \sigma)= \{\atts' | \F' =(\args,\atts'), T \subseteq (E \in \sigma(\F'))\}$).
\end{definition}

\begin{table}[t]
	\centering
	\begin{tabular}{ccc}
		$\sigma$ &  strict & non-strict  \\
		Admissible & P  & \bf{NP-c}   \\
		Complete & \bf{NP-c} &  \bf{NP-c}  \\
		Preferred & $\sum^{P}_{2}$-c  &  \bf{NP-c}  \\
		Stable &  P & \bf{NP-c}  \\
		Grounded & \bf{NP-c} & \bf{NP-c}
	\end{tabular}
        \captionsetup{width=3.1\linewidth}
	\caption{The complexity of extension enforcement~\cite{extenf}; in bold we highlight NP-Complete problems.}\label{tab:complexity2}
\end{table}

If we still consider the framework $\F$ in Figure~\ref{fig:argnetex}, we have for example that if we want to strictly enforce $T = \{a, e\}$ as a complete extension, the solution corresponds to $\{a \atts b, c\atts b, c \atts d, d \atts c,\}$: it is enough to remove $d \atts e$ from the attacks.

\subsection{Quantum Annealing and QUBO}\label{sec:qubo}

Quantum Annealing is a procedure \cite{quantanneal} that employs a quantum computing device to solve optimization problems formulated in terms of finding a configuration of minimum energy.

It is based on the Quantum Adiabatic Theorem and it employs quantum tunneling effects to produce optimal or near/optimal solutions of a discrete optimization problem. 

The process starts with an initial function $\phi_s$, for which the solution that minimizes the energy is easy to find, and slowly transitions to the final function $\phi_f$, which corresponds to the function to be optimized. If the transition is enough slow, the \emph{Quantum Adiabatic Theorem} guarantees that the solution with the minimum energy adapts to the change of the objective function.

D-Wave produces a series of quantum annealers, 
which are computing devices implementing the quantum annealing
algorithm. In particular, the most advanced machines are able to handle problems with thousands of variables.

Quantum annealers can be "programmed" by describing the problems as \emph{Quadratic Unconstrained Binary Optimization} (in short, \emph{QUBO}) or as Ising models \cite{ising}.

In this paper, we employ the QUBO formalism. It is expressive enough to encode several optimization problems formulated in various application domains ~\cite{qubosurvey}.


QUBO has been intensively investigated and is used to characterize and solve a wide range of optimization problems: for example, it encompasses SAT Problems, Constraint Satisfaction Problems, Maximum Cut Problems, Graph Coloring Problems, Maximum Clique Problems, General 0/1 Programming Problems, and many more~\cite{qubomodel}. Moreover, QUBO embeddings exist for Support Vector Machines, Clustering algorithms, and Markov Random Fields~\cite{quboml}.

In a QUBO problem, $n$ binary variables $x_1, \dots, x_n$ and an $n \times n$ upper-triangular matrix $Q$ are used to formulate the task, which involves minimizing (or maximizing) the function

$$f(x) = \sum_{i=1}^n Q_{i,i} x_i + \sum_{i < j}^n Q_{i,j} x_i x_j$$

The diagonal terms $Q_{i,i}$ are the linear coefficients, and the non-zero off-diagonal terms $Q_{i,j}$ are the quadratic coefficients. This can be expressed more concisely as

$$\min_{x \in \{0,1\}^n} x^T Q x$$

\noindent where $x^T$ denotes the transpose of the vector $x$. The square matrix of coefficients can be organized in a symmetric way, where for all $i, j$ except $i=j$, $Q_{i,j}$ is replaced by $(Q_{i,j} + Q_{j,i}) / 2$, or, as stated before, in an upper-diagonal form where for all $i, j$ s.t. $i>j$, $Q_{i,j}$ is replaced by $Q_{i,j} + Q_{j,i}$ and then all $Q_{i,j}$ are replaced by $0$ for $j < i$.

The formulation of a discrete constrained optimization problem as QUBO requires the following steps: \emph{i)} find a binary representation for the solutions, and \emph{ii)} define a penalization function that penalizes unfeasible solutions (i.e., violating a constraint).

Except for the 0/1 limitations on the decision variables, QUBO is an unconstrained model in which the Q matrix contains all the problem data. Due to these features, the QUBO model presents an innovative perspective on classically limited representations and is especially appealing as a modeling framework for combinatorial optimization issues. Rather than applying constraints in the conventional sense, classical constrained models can be successfully reformulated as QUBO models by inserting quadratic penalties into the objective function. Minimization problems are enhanced by applying penalties to produce an augmented objective function, which must be minimized. The augmented objective function becomes equivalent to the original function if the penalty terms are reduced to zero.

When optimizing a problem with constraints, choosing a penalty that is too small can result in unfeasible solutions. However, selecting a large penalty value to enforce the constraints can cause difficulties for the optimization algorithm in finding feasible solutions. This approach can also lead to other issues, such as sensitivity to penalty values, increased computational efforts during optimization, and instability during the solver's iterations. For this reason, a sensitive amount of literature has been produced to find good coefficients~\cite{optweights1,optweights2}, or techniques to model classical constraints into a function, e.g., a logical \emph{and} constraint between $x_1$ and $x_2$ as a multiplication $x_1 \cdot x_2$.

The literature discussing precise approaches for QUBO using conventional computers consists of various algorithms~\cite{qubosurvey}, all of which share the characteristic that, if given enough time and memory, they end with a globally optimal solution. Most of these techniques use a standard branch-and-bound tree search, although alternative methods are also available. For example, in \cite{exact1}, a QUBO solution is based on the inherent geometric properties of the minimum circumscribed sphere that contains the ellipsoidal contour of the objective function, while in \cite{exact2}, the authors adopt Lagrangean decompositions.

Many research papers have been published in recent years due to the NP-hardness and potential applications of QUBO. These papers describe various heuristic approaches to quickly find high-quality solutions for medium- to large-sized problem cases. Although some of these techniques are simple enough to be called heuristics, the most effective are metaheuristic processes that use compound strategies that are more advanced than those found in basic heuristics. For example, in addition to simulated annealing~\cite{simannealqubo}, the work in \cite{tabu} presents a guided tabu-search algorithm alternating between a basic tabu-search procedure and a variable fix/free phase. In contrast, \cite{metaqubo}  presents a hybrid metaheuristic approach that adopts crossover and update operators and then tabu-search to examine the solutions of the offspring.

\subsection{Simulated Annealing}\label{sec:simulated}
\emph{Simulated Annealing} (\emph{SA}) is a probabilistic optimization algorithm inspired by the process of annealing in metallurgy, where a material is heated and then slowly cooled to reach a low-energy state~\cite{sannealing}. Three main steps characterize the algorithm: \emph{i)} initialization, \emph{ii)} temperature initialization, and \emph{iii)} iteration. 

With initialization, the algorithm starts with an initial solution to the optimization problem; this solution can be generated randomly or by using some heuristic method. In temperature initialization, an initial temperature $T$ is set; $T$ determines the probability of initially accepting worse solutions. The temperature gradually decreases over time. During the iteration phase, the algorithm repeats until a stopping criterion is met (e.g., a maximum number of iterations or when the temperature drops below a certain threshold).

During iteration, the first step usually consists of perturbing the current solution to generate a neighboring solution; this perturbation can involve minor changes to the current solution. Afterward, the algorithm calculates the corresponding change $\Delta(\mathit{Energy})$ in the value of the objective function between the current solution and the neighboring solution.
If $\Delta(\mathit{Energy})$ is negative (i.e., the neighboring solution is better), the algorithm accepts the neighboring solution. Otherwise,
the algorithm accepts the neighboring solution with a probability determined by a temperature-dependent acceptance probability function. This allows the algorithm to escape local minima and explore the solution space more effectively. In the final step of the iteration phase, the algorithm updates the current solution if the neighboring solution is accepted, and it decreases the temperature according to a cooling schedule, which controls the rate at which the temperature decreases. The cooling schedule can be linear, exponential, or follow other schemes.
The algorithm stops when the stopping criterion is met, and the best solution found during the iterations is finally returned.

Simulated Annealing allows the algorithm to explore the solution space by initially accepting worse solutions with a certain probability, which gradually decreases as the temperature decreases. This property enables the algorithm to escape local optima and potentially find better solutions. The cooling schedule balances exploration and exploitation, determining how quickly the algorithm converges to an optimal solution.

\section{Encoding}\label{sec:encoding}

This section presents a QUBO encoding of some Argumentation problems in Section~\ref{sec:background}. In Section~\ref{sect:encodingclassic}, we focus on all the NP-Complete cases shown in Table~\ref{sec:complexity} because solving QUBO is NP-Complete in turn. More specifically, the encoded problems are  
$\dc\textit{-}\sigma$, 
$\ex\textit{-}\sigma$, and 
$\nem\textit{-}\sigma$, 
while the considered semantics are $\sigma = \{\ad, \co, \pr, \st, \sst\}$.

After that, in Section~\ref{sect:enforcenc}, we formulate an encoding for extension enforcement problems as introduced in Section~\ref{sect:bgenforcement}; in particular, we focus on all NP-Complete problems regarding strict enforcement in Table~\ref{tab:complexity2} (i.e., with complete and grounded semantics).

\subsection{Encoding Extension-based Semantics Problems}\label{sect:encodingclassic}

The encoding of problems in Argumentation uses a set of $n$ binary variables $x_1,\dots,x_n$ associated with the arguments $\{a_1,\dots,$ $a_n\}$ in $\args$. 
The variables $x_1,\dots,x_n$ represent a subset $E$ of $\args$: $a_i\in E$ if and only if $x_i=1$. We denote by $\underbar{x}$ the tuple of variables $(x_1,\dots,x_n)$ and by $\mathbf{x}\in\{0,1\}^n$ a vector of possible values for $x_1,\dots,x_n$. Each semantics $\sigma$ will be associated with a quadratic penalty function $P_{\sigma}$ such that $P_{\sigma}$ assumes its minimum value in $\mathbf{x}$ if and only if the corresponding set $E=\{a_i\in \args \,:\, x_i=1\}$ is an extension valid for $\sigma$. 

Most of the argumentation semantics require admissible sets. Hence, we define a penalty function $P_{adm}$ that enforces this property, which is the sum of 4 terms. 
The first term forces the set $E$ to be \textbf{conflict-free}:
\begin{equation}
P_{\mathit{cf}}=\sum_{i \atts j} x_i x_j
\end{equation}

The value of $P_{cf}$ corresponds to the number of internal attacks in $E$, and its value is $0$ if and only if $E$ is conflict-free. 

The constraints to model the notion of \textbf{defense} are more complicated and require some sets of additional variables. The first set contains the variables $t_1,\dots,t_n$, denoting which arguments are attacked by $E$: $t_i=1$ if and only if some argument of $E$ attacks $a_i$. 

The variables $d_1,\dots,d_n$ of the second set denote which arguments are defended by $E$: $d_i=1$ if and only if $a_i$ is defended (from all possible attacks) by some arguments of $E$. For each argument $a_i$, the penalty function $P_{t}^i$ forces $t_i$ to be $1$ if and only if $a_i$ is attacked by $E$, i.e., $t_i = \bigvee_{j \atts i} x_j$.



The function
\begin{equation}
OR(z,x,y)=z+x+y+xy-2z(x+y)
\end{equation} 

is used to express the constraint that the binary variable $z$ is the disjunction $z=(x\mbox{ or }y)$ of the binary variables $x,y$ as a quadratic function, as shown in \cite{rosenberg}. 

Each penalty function $P_{t}^i$ requires $\max\{h_i-2,0\}$
auxiliary variables, where $h_i$ is the number of attackers of $a_i$. Of course, if $h_i\le 2$, no additional variable is required. More details can be found in \cite{pricai22}.




The other penalty function $P_{d}^i$ forces $d_i$ to be $1$ if and only if $a_i$ is defended by $E$, i.e.,
$ d_i = \bigwedge_{j \atts i} t_j $. The term is encoded by means of the function \cite{rosenberg}
\begin{equation}
AND(z,x,y)=3z+xy-2z(x+y)
\end{equation}
which expresses the conjunction $z=(x\mbox{ and }y)$ of binary variables $x,y$ as a quadratic function.







In addition, $P_{d}^i$ requires $\max\{h_i-2,0\}$ auxiliary variables. The final term 

\begin{equation}
P_{def}= \sum_{i=1}^n x_i(1-d_i) 
\end{equation}

\noindent forces each argument in $E$ to be defended by $E$. Summing up, the penalty function for \textbf{admissible} sets is 
\begin{equation}
P_{adm}=P_{cf}+ \sum_{i=1}^n P_t^i + \sum_{i=1}^n P_d^i  + P_{def}
\end{equation}

It is easy to prove that the minimum value of $P_{adm}$ is $0$, and the related values for $\underbar{x}$ correspond to admissible sets.

It is important to note that the total number of binary variables needed for $P_{adm}$ is
$$ N = 3n+2\sum_{i=1}^n \max(h_i-2,0).$$

Note that if $h=\max (h_i)$, then $N=O(nh)$. For the \textbf{complete} semantics, we  need to add an additional term to $P_{adm}$ which forces all the arguments defended by $E$ to be elements of $E$: 

\begin{equation}
P_{co}=P_{adm} + \sum_{i=1}^n (1-x_i)d_i
\end{equation}

Concerning the \textbf{preferred} semantics, since the problem of verifying whether a set $E$ belongs to $\pr$ is  \textit{coNP}-Complete, it is unlikely to find a formulation of this semantics in QUBO with a polynomial number of binary variables.

However, it is possible to find preferred extensions of maximum cardinality with QUBO by minimizing the combination of the penalty function $P_{co}$ and a term that corresponds to $|E^C|$, i.e., the size of the complementary of $E$: 

\begin{equation}
P_{pr}=\sum_{i=1}^n (1-x_i)+(n+1)\cdot P_{co}
\end{equation}

It is easy to see that the minimum value of $P_{pr}$ is $s\le n$, and the related values for $\underbar{x}$ correspond to complete sets with maximum cardinality $s$. On the other hand, any combination of values $\underbar{x}$ such that $P_{pr}$ is greater than $n$ corresponds to a set of arguments that is not complete. It is essential to understand that the extensions found with this method can only be a subset of all $\pr$ extensions.

The considerations about the \textbf{semi-stable} semantics are similar to the preferred semantics because of coNP-Completeness of the verification problem. Again, it is possible to find a semi-stable extension of maximum cardinality by minimizing the following objective function:

\begin{equation}
P_{sst}= \sum_{i=1}^n (1-t_i) + (n+1) P_{co}
\end{equation}

\noindent where the additional term corresponding to the number of arguments not attacked is added to $P_{co}$. Hence, the minimum value of $P_{sst}$ is $s\le n$, and the related values of $\underbar{x}$ correspond to complete sets where the number of arguments not attacked is $s$. Even in this case, the method is not complete because it cannot find semi-stable extensions whose cardinality is less than $s$.


Finally, the encoding of the \textbf{stable} semantics is obtained by adding an additional term to $P_{co}$ which forces all the arguments not belonging to $E$ to be attacked by $E$:

\begin{equation}
P_{st}=P_{co}+ \sum_{i=1}^n (1-x_i)(1-t_i)
\end{equation}

It is straightforward to prove that the minimum value of $P_{st}$ is $0$  if and only if the corresponding $E$ belongs to $\st$. Hence, this encoding can solve the task $\ex\textit{-}\st$.

%
%
%
%


To express that $E$ is not empty (that is, solving task $\nem$) in a given semantics $\sigma$, it is sufficient to add to the corresponding penalty function $P_{\sigma}$ a term which enforces the constraint


\begin{equation}\label{nonempty}
  \bigvee\limits_{i=1}^n x_i = 1.  
\end{equation}

In general, the minimum value of $P_{\sigma}+P_{ne}$ is $0$
if and only if $E$ is not empty.

This task can also be solved for $\pr$ and for $\sst$ semantics using QUBO because, although there are no complete encodings for $\pr$ and for $\sst$, the existence of a non-empty extension for these semantics is reduced to the same problem for $\ad$ semantics~\cite{dvorakcomplexity}.

It is important to notice that the number of binary variables needed to solve the task $\nem$ increases to $$ N+n-2=4n-2+2\sum_{i=1}^n \max(h_i-2,0)$$ because the term (\ref{nonempty}) requires $n-2$ additional variables.


To express that a given argument $a_i$ must appear in $E$ (i.e., the $\dc$ task), it is sufficient to force $x_i$ to be $1$ and propagate this setting in all encodings, thus obtaining a simplified quadratic function, with a reduced number of binary variables. It is easy to see that the minimum value of this function is $0$ if and only if
$a_i$ is credulously accepted.

Furthermore, QUBO can solve the credulous acceptance for $\pr$ because this problem is equivalent to checking the credulous acceptance in admissible sets~\cite{dvorakcomplexity}. In Section~\ref{sec:comparison}, we will use such an encoding during tests.

Finally, it is possible to use QUBO to solve the negative formulation of the skeptical acceptance (the $\sc$ task), i.e., to verify that a given argument $a_i$ is not contained in all $E \in \sigma(\F)$. 
It is sufficient to replace $x_i$ with $0$ and propagate this setting in the encoding of $\sigma$. It is easy to see that the minimum value of this function is $0$ if and only if $a_i$ is not skeptically accepted.

To clarify the presented encoding, we introduce Example~\ref{ex:encoding} where $\dc$-$\co$ and $\dc$-$\st$ problems are presented.

\begin{example}\label{ex:encoding}
Taking into account the framework in Figure~\ref{fig:argnetex}, we denote the arguments with the indices $a_1=a, a_2=b, a_3=c, a_4=d,$ and $a_5=e$.
	
In the first step, we obtain $P_{cf}=x_1 x_2+x_2 x_3+x_3 x_4+x_4 x_5$. The defense encoding does not need any auxiliary variables other than $t_i$ and $d_i$ because each argument has at most two attackers. The variables $t_i$ are constrained as $$t_1=0,\, t_2=x_1 \vee x_3, \, t_3=x_4, \, t_4=x_3, \, t_5=x_4.$$ Therefore the only penalty term in $P_{adm}$ is related to $t_2$. 
After some simplification, the variables $d_i$ are constrained as
$$d_1=1,\, d_2=0, \,d_3=x_3, \,d_4=x_4, \,d_5=x_3.$$
	
The penalty function $P_{def}=x_1(1-d_1)+x_2(1-d_2)+x_3(1-d_3)+x_4(1-d_4)+x_5(1-d_5)$ is simply $P_{def}=x_2+x_5(1-x_3)$ because, for any binary variable $x_i$, the term $x_i(1-x_i)$ reduces to $0$. Therefore,  
\begin{equation}
P_{adm}=x_1 x_2+x_2 x_3 + x_3 x_4 + x_4 x_5 + x_2 + x_5 (1-x_3)
\end{equation}

Note the term about $t_2$ 
(i.e., $OR(t_2,x_1,x_3)$) 
can be ignored because $t_2$ does not appear in the rest of the formula. The penalty function 
$$P_{co}= P_{adm}+(1-x_1)d_1+(1-x_2)d_2+(1-x_3)d_3+(1-x_4)d_4+(1-x_5)d_5$$

reduces to 
\begin{equation}
P_{co}=x_1 x_2+x_2 x_3 + x_3 x_4 + x_4 x_5 + x_2 + x_5 + x_3 -2x_3x_5 + 1-x_1
\end{equation}
	
This function is the starting point for solving the task $\dc$-$\co$:, given any argument $a_i$, it is sufficient to set $x_i$ to $1$, simplify $P_{co}$, and minimize it.
Regarding the $\st$ semantics, the objective function is
\begin{equation}
P_{st} = P_{co}+(1-x_1)(1-t_1)+(1-x_2)(1-t_2)+ 2(1-x_3)(1-x_4)+(1-x_5)(1-x_4)
\end{equation}

\noindent which finally reduces to
\begin{multline}
P_{st} =  x_1 x_2 + x_2 x_3 + 3x_3 x_4 + 2x_4 x_5 -x_3 -3x_4 - 2x_3 x_5 + 
2(1-x_1) - t_2 + x_2 t_2 +x_4 x_5 + 5 + \\ AND(t_2, x_1,x_3) 
\end{multline}

The term concerning $t_2$ cannot be ignored because $t_2$ occurs in the objective function.
\end{example}

\subsection{Extension Enforcement in QUBO}\label{sect:enforcenc}
An important subject that has emerged in the literature in recent years is dynamic changes in AFs, as introduced in Section~\ref{sect:enforcenc}. In particular, attention has been paid to the problem of enforcing a set $E$ of arguments, i.e., ensuring that $E$ is an extension (or a subset of an extension) of a given framework $\F$.

The extension enforcement task can be formulated with similar techniques.
Let us focus on the strict version of this problem concerning complete semantics (see Table~\ref{tab:complexity2}).  To simplify the notation, the arguments in the set $T$ are the first $k$ arguments $a_1,\dots,a_k$ in $\args$.

We use a first set of binary variables $r_{ij}$, for $i,j=1,\dots,n$. Each variable $r_{ij}$ is $1$ whether in the new attack relationship $\atts'$, $a_i$ attacks $a_j$. Moreover, we use binary variables $t_i$, for $i=1,\dots,n$, as in the encoding of credulous acceptance in Section~\ref{sect:encodingclassic}. Finally, a new set of binary variables $s_{ji}$, for $j=k+1,\dots,n$ and $i=1,\dots,k$ is used.

We define a penalty function $\tilde{P}_{co}$, which is zero if and only if $T$ is a complete extension under the attack relationship described by $r_{ij}$. $\tilde{P}_{co}$ is the sum of $4$ terms.

Since the completeness of $T$ requires $\atts'$ to be conflict-free, all the values of $r_{ij}$, when $1\le i,j \le k$, are set to $0$. In this way, there are no possible internal attacks in $T$. 

The first term of $\tilde{P}_{co}$ is 
\begin{equation}
\tilde{P}_s=\sum_{j=1}^n\sum_{i=1}^n AND(s_{ji}, r_{ji}, 1-t_j)
\end{equation}

\noindent which forces the values of each variable $s_{ji}$ to correspond to the product $r_{ji}(1-t_j)$. Note that $s_{ji}$ is $1$ if and only if $r_{ji}=1$, i.e., if $a_j$ attacks $a_i$, and $t_j=0$, i.e., if $a_j$ is not counterattacked by $E$. 

The second term is 
\begin{equation}
\tilde{P}_t=\sum_{i=1}^n \tilde{P}^{i}_{t}
\end{equation}

\noindent where  $\tilde{P}^{i}_{t}$, for each $i=1,\dots,n$, enforces the constraint $t_i=\bigvee_{j=1}^k r_{ji} $, which means that $t_i=1$ if and only if the argument $a_i$ is attacked by some argument $a_j\in T$. This term is encoded in QUBO using $k-2$ auxiliary binary variables, similar to that used for $P^i_t$.

The third term is 

\begin{equation}
\tilde{P}_{ad}=\sum_{i=1}^k \sum_{j=k+1}^n s_{ji}
\end{equation}

\noindent which adds a penalty each time an attacker $a_j$ of some argument $a_i\in  T$ is not counterattacked by $T$. It is then easy to see that $\tilde{P}_s+\tilde{P}_t+\tilde{P}_{ad}=0$ if and only if $T$ is an admissible extension concerning $\atts'$.

The last term is 

\begin{equation}
\tilde{P}_{nd}=\sum_{i=k+1}^n \tilde{P}^i_{nd}
\end{equation}

\noindent where $\tilde{P}^i_{nd}$, for each $i=k+1,\dots,n$, enforces the constraint $\bigvee_{j=1}^n s_{ji} = \text{true}$. This constraint means that the argument $a_i \not\in T$ is not defended by $T$ against some attackers. The term $\tilde{P}^i_{nd}$ is encoded in QUBO using a new set of $n-2$ auxiliary variables to represent that the disjunction of the variables $s_{ji}$ is true. Namely, 
\begin{multline}
    \tilde{P}^i_{nd} =  (1-s_{1,i})(1-\sigma_{1,i}) +OR(\sigma_{1,i},s_{2,i},\sigma_{2,i}) +\dotsc  + \\ OR(\sigma_{n-3,i},s_{n-2,i},\sigma_{n-2,i}) +
 OR(\sigma_{n-2,i}, s_{n-1,i}, s_{n,i})
\end{multline}

\noindent and from this, $\tilde{P}_{co}=\tilde{P}_{s}+\tilde{P}_{t}+\tilde{P}_{ad}+\tilde{P}_{nd}$. The overall objective function to be minimized is 

\begin{equation}
    f_{co}=\sum_{a_i \atts a_j} (1-r_{ij}) + \sum_{a_i \not\atts a_j} r_{ij} + \lambda \tilde{P}_{co},
\end{equation}

\noindent where $\lambda$ is a constant, whose value should be large enough such that the minimum of $f_{co}$ is obtained when $\tilde{P}_{co}=0$.

It is easy to see that the number of binary variables needed to encode the enforcement task is 
$$ 3n^2+n(k-3)$$
which is quadratic with respect to the number $n$ of arguments.



\section{Implementation and Tests on Conventional Machines}\label{sec:implementation}

This section presents tests to empirically validate our QUBO encoding of NP-Complete problems on AFs, and to measure performance by also making a comparison with related approximate solvers where possible: for problems concerning the extension enforcement, it was not possible to find publicly released solvers that work on this problem, and so Section~\ref{sect:testextenforc} present stand-alone tests. On the other hand, the next section (Section~\ref{sec:comparison}) also compares the other two tools that solve the same problems. The experiments in this section are executed on a personal laptop with 2 GHz Quad-Core Intel Core i5 and a RAM of 16 GB running at 3733 MHz.

\subsection{Tests and Comparison on Classic Problems}\label{sec:comparison}

The primary goal of the tests in this section is to prove the correctness of the encodings presented in Section~\ref{sect:encodingclassic} (referred to as \emph{QArg} in the following) for all the NP-Complete problems in Table~\ref{sec:complexity}: to do this, we compared all the results obtained by approximate solvers with \emph{ConArg}~\cite{conarg,conarg2}, a reduction-based  (exact) solver using \emph{Constraint Programming}, which has also been used as the reference solver of ICCMA19. As a secondary goal, we compared the results of QArg considering the other two approximate solvers overviewed in Section~\ref{sec:related}, i.e., Harper++ and AFGCN. Harper++, developed by M. Thimm, was downloaded from its Website\footnote{Tweety Abstract Argumentation Solvers: \url{http://taas.tweetyproject.org}.}, while AFGCN was provided directly by L. Malmqvist, i.e., the author himself. To our knowledge, this section consists of the first comparison among approximate solvers in addition to ICCMA21.

All the tests in this section were performed on a 2GHz Quad-Core Intel Core i5 with 16 GB of RAM. We set a timeout of $60$ seconds, as applied in ICCMA21 when considering approximate solvers (instead of $600$ seconds when considering exact solvers, still in ICCMA21). 

All QArg tests were run locally using the SA algorithm provided by the Ocean SDK package in a Python \emph{SimulatedAnnealingSampler} class. For all experiments, we set the parameter \emph{number of reads} of the algorithm to $\mathit{nArguments} \times 2$ ($\mathit{nArguments}$ is the number of arguments in the considered instance): each read is generated by one run of the
SA algorithm. We set \emph{number of sweeps} used in annealing to $\min(\mathit{nArguments}\times 50, 1000)$. In case no solution with energy $0$ is found after a run,  the initial random \emph{seed} is changed, and a successive iteration of SA is executed until a zero-energy solution is found or the timeout is met, with a limit of $100$ iterations for each framework.

We adopted four different benchmarks, called \BA, \BB,  \BC, and \BD, respectively.
\BA consists of $100$ \emph{``Small''}  instances that are part of the benchmark selected for ICCMA19. These instances have $5$ to $191$ arguments (median $28.5$) and $8$ to $8192$ attacks (median $296$).  Moreover, we considered the same arguments as used in ICCMA19 to check credulous acceptance. \BA consists of a practically-oriented benchmark on logic-based AF instantiated from inconsistent knowledge bases expressed using Datalog$^\pm$, a language widely used in Semantic Web.\footnote{\emph{Benchmark on Logic-Based Argumentation Framework with Datalog$^\pm$}, by Bruno Yun and Madalina Croitoru: \url{http://argumentationcompetition.org/2019/papers/ICCMA19_paper_2.pdf}.}

Since from internal tests, we noticed that all the possible arguments in those frameworks are credulously accepted ($\dc$-$\co$ always returns $100$ ``YES''), to test ``YES' answers, we considered a second benchmark, 
that is \BB. It contains $100$ frameworks with $80$ arguments each, generated as \emph{Erd\H{o}s-R\'{e}nyi} graphs~\cite{erdos}. The generated instances have $271$ to $436$ attacks (median $349$).
In the ER model, a graph is constructed by randomly connecting $n$ nodes. Each edge is included in the graph with probability \emph{p} independent of every other edge. Clearly, as $p$ increases from $0$ to $1$, the model becomes more and more likely to include graphs with more edges. To create our benchmark, we adopt $p=c \cdot log (n) / n$ (with $n$ the number of nodes and $c$ empirically set to $2.5$), ensuring the connectedness of these graphs. The ER model has also been used as part of the benchmark for all ICCMA competitions since ICCMA15. For each of these $100$ frameworks, we randomly selected one argument that is not credulously accepted for the considered semantics; we checked the acceptance of the argument using ConArg, which is an exact solver.

\begin{figure*}[t]
	\centering
	\begin{subfigure}{0.32\linewidth}
		\includegraphics[width=1\linewidth]{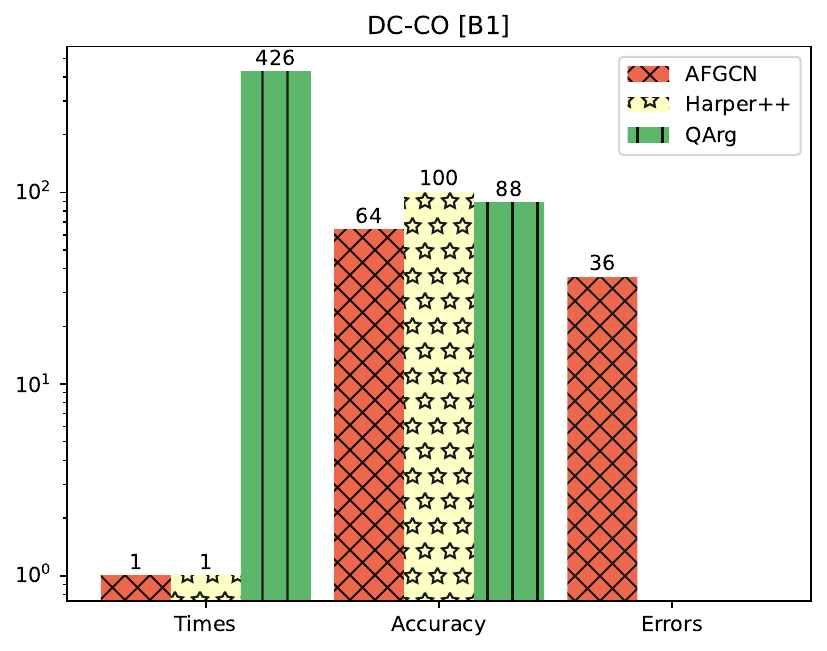}
		\label{fig:dcco}
	\end{subfigure}
	\begin{subfigure}{0.32\linewidth}
		\includegraphics[width=1\linewidth]{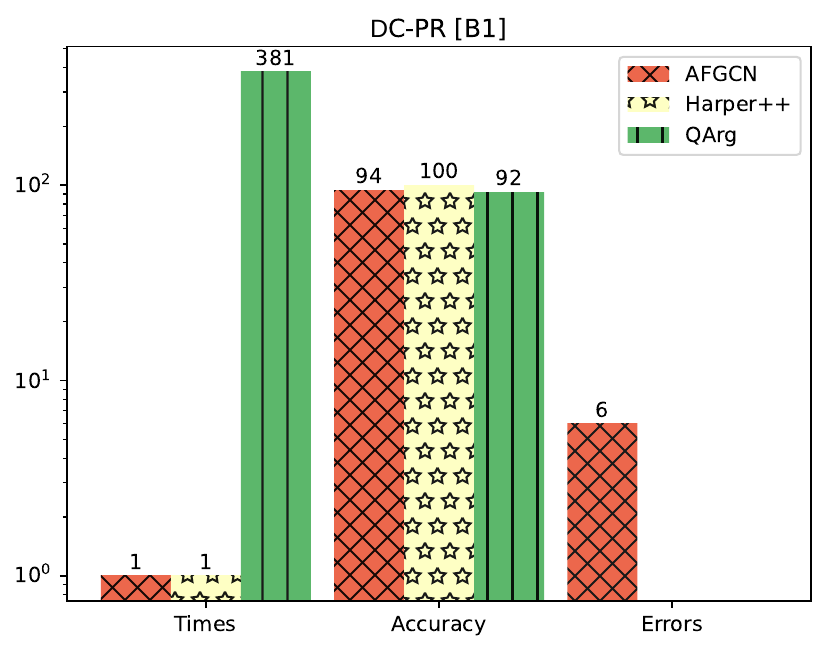}
		\label{fig:dcpr}
	\end{subfigure}
	\begin{subfigure}{0.32\linewidth}
		\includegraphics[width=1\linewidth]{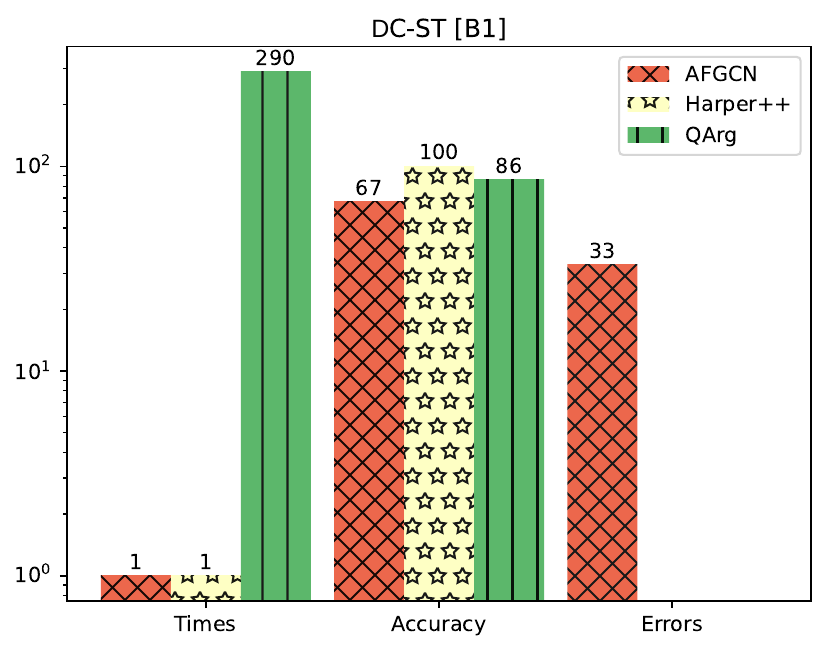}
		\label{fig:dcst}
	\end{subfigure}
	\caption{The $\dc$-$\co$, $\dc$-$\pr$ and $\dc$-$\st$ tested on \BA ($100$ ``YES'') by comparing AFGCN, Harper++, and our QUBO encoding (QArg). Considering QArg, the difference between $100$ and ``Accuracy'' is the number of timeouts.}\label{fig:dcyes}
\end{figure*}

\begin{figure*}[t]
	\centering
	\begin{subfigure}{0.32\linewidth}
		\includegraphics[width=1\linewidth]{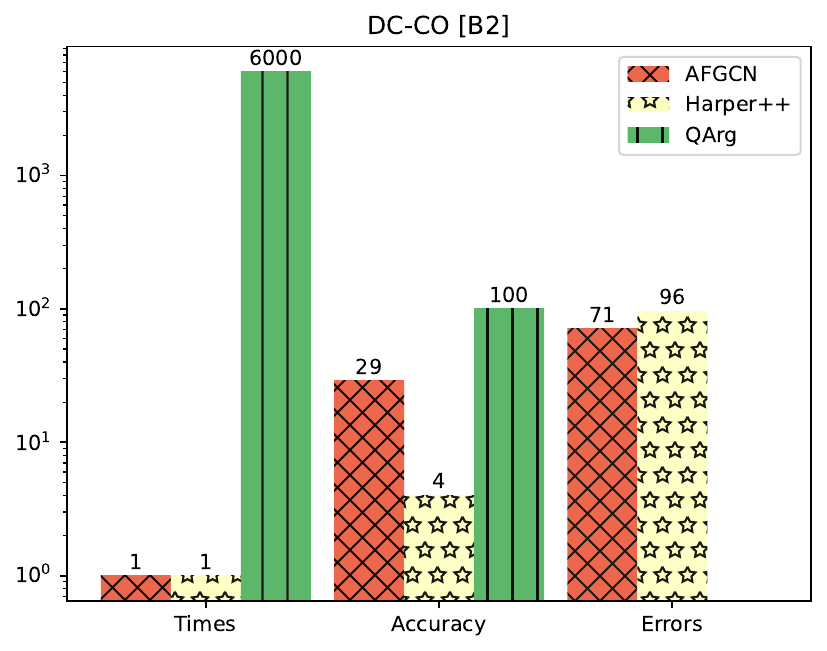}
		\label{fig:dccob2}
	\end{subfigure}
	\begin{subfigure}{0.32\linewidth}
		\includegraphics[width=1\linewidth]{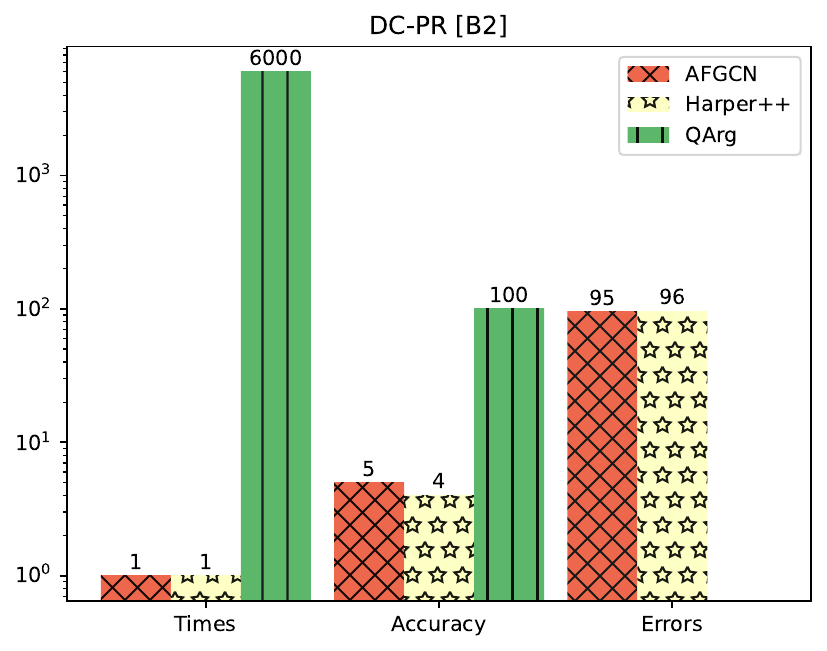}
		\label{fig:dcprb2}
	\end{subfigure}
	\begin{subfigure}{0.32\linewidth}
		\includegraphics[width=1\linewidth]{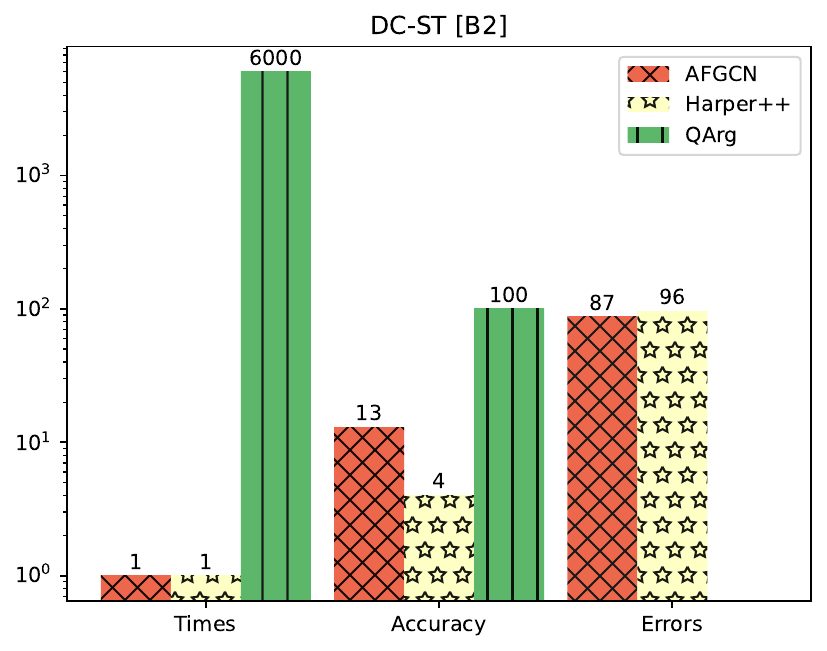}
		\label{fig:dcstb2}
	\end{subfigure}
	\caption{The $\dc$-$\co$, $\dc$-$\pr$ and $\dc$-$\st$ tested on \BB ($100$ ``NO'') by comparing AFGCN, Harper++, and our QUBO encoding (QArg). Considering QArg, the difference between $100$ and ``Accuracy'' is the number of timeouts.}\label{fig:dcno}
\end{figure*}



\begin{figure*}[t]
\centering
	\begin{subfigure}[t]{0.47\linewidth}
		\includegraphics[width=0.9\linewidth]{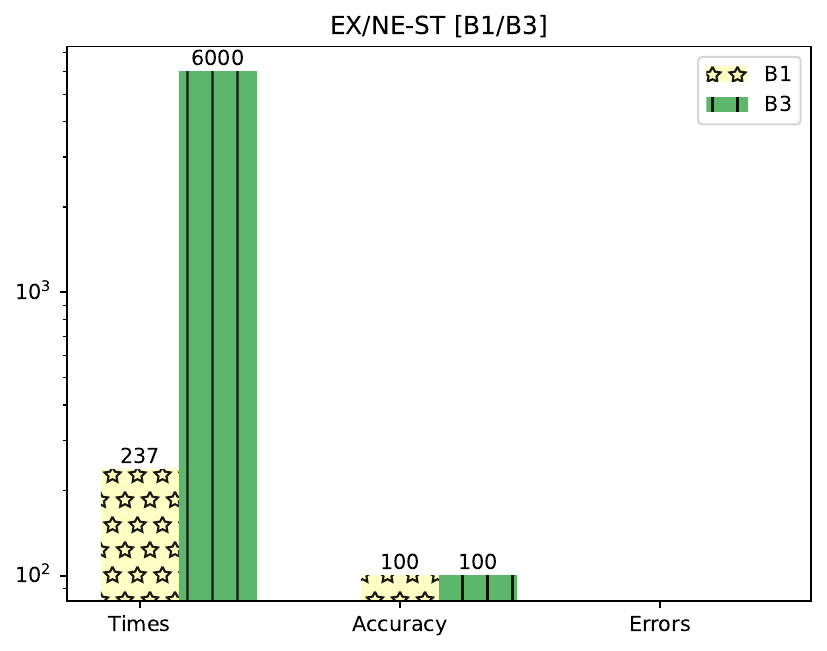}
  \caption{The $\ex$-$\st$ and $\nem$-$\st$ tasks, which reduces to the same problem~\cite{dvorakcomplexity}, are tested on $\BA$ ($100$ ``YES'') and $\BC$ ($100$ ``NO'').  AFGCN and Harper++ do not solve these tasks.}
		\label{fig:dccob2}
	\end{subfigure}
 \hspace{0.3cm}
	\begin{subfigure}[t]{0.47\linewidth}
		\includegraphics[width=0.9\linewidth]{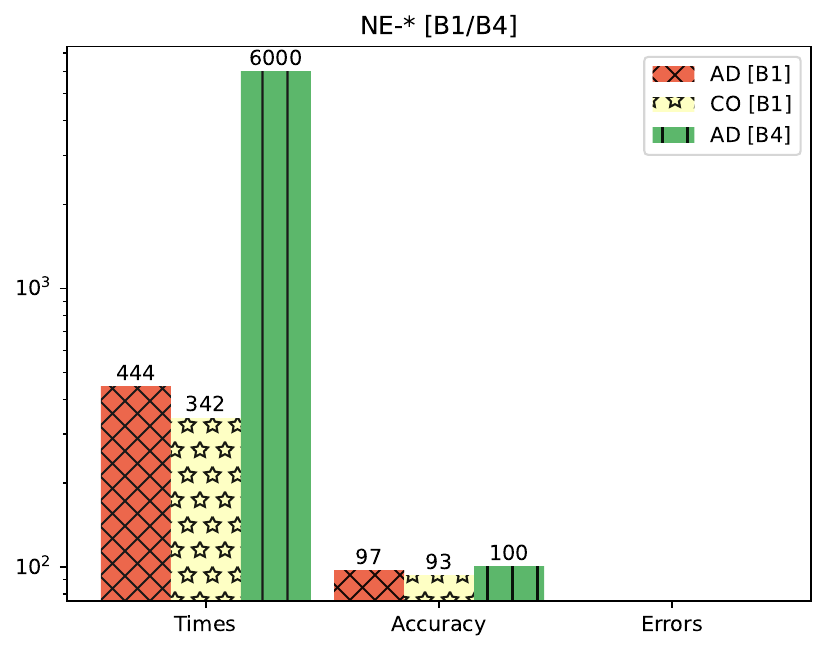}
  \caption{The $\nem$-$\ad$ and $\nem$-$\co$ tasks tested on \BA ($100$ ``YES''), and $\nem$-$\co$ tasks tested on \BD ($100$ ``NO''). They also count as $\nem$-$\pr$ and $\nem$-$\sst$~\cite{dvorakcomplexity}. AFGCN and Harper++ do not solve these tasks. The difference between $100$ and ``Accuracy'' is the number of timeouts.}\label{fig:dcprb2}
	\end{subfigure}
	\caption{}\label{fig:dccob2dcprb2}
\end{figure*}

The benchmark \BC was created to test the non-existence of a stable extension, that is, $\ex$-$\st$ on frameworks where no such an extension exists, i.e.,  frameworks returning ``NO'' as the correct answer. To our knowledge, all previous editions of the ICCMA competition do not include any established benchmark in this sense since $\ex$ is not among the tasks tested in the competition ($\se$ is usually tested instead. For this reason, we created $\BC$ by adding some new arguments to each framework of \BA to introduce an odd-length cycle of length between $1$ and $7$ (randomly) and being odd-cycle-free guarantees at least one stable extension~\cite{Dung:1995}. \BC is thus formed of $100$ frameworks with no stable extension (checked with ConArg).

Finally, we created \BD to test the absence of a non-empty extension while testing $\nem$ in $\ad$, $\co$, $\pr$, and $\sst$, which are an NP-Complete problems as well (see Table~\ref{sec:complexity}). We created this dataset using all the $100$ frameworks in $\BA$. We added a self-attack to all the arguments that appear in an admissible extension for each framework. This caused all frameworks to have only $\emptyset$ as admissible, complete, preferred, and semi-stable extensions. Therefore, testing $\nem$ on \BD should lead to $100$  ``NO'' answers.

Figure~\ref{fig:dcyes} shows the comparison while testing $\dc$-$\co$ on $\BA$ ($100$ ``YES''). The \emph{Times} bars report the cumulative time in seconds to solve all the frameworks, excluding timeouts; \emph{Accuracy} bars show the number of instances correctly solved before the timeout, while \emph{Errors} bars show the number of instances for which the answer differs from the reference (exact) solver,\footnote{No bar in a figure means $0$ errors.} i.e., ConArg.

QArg has $12$, $8$, and $14$ timeouts in $\dc$-$\co$, $\dc$-$\pr$, and $\dc$-$\st$; these can be considered as errors if QArg returns ``NO'' if no solution is found before the timeout. In this experiment, AFGCN reports a high number of errors in the case of complete and stable semantics. As introduced in Section~\ref{sec:encoding}, we use the encoding for $\dc$-$\ad$ to test $\dc$-$\pr$, since the two problems are equivalent~\cite{dvorakcomplexity}.

Figure~\ref{fig:dcno} presents the same comparison testing $\dc$-$\co$ on $\BB$ ($100$ ``NO''). In this case, QArg always meets the timeout of $60$ seconds because all arguments are not credulously accepted. By returning ``NO'' after the timeout, QArg always provides the correct answer, while the other two solvers both return a very high number of errors: they answer ``YES''.

For the remaining tests, both Harper++ and AFGCN cannot solve the following tasks (see Section~\ref{sec:related}), and for this reason, we will report the results only for QArg.

Looking at Table~\ref{sec:complexity}, the problem $\ex$ is hard for the stable semantics only, as all the other semantics always lead to at least one extension. Moreover, in frameworks with at least one argument, the problems $\ex$-$\st$ and $\nem$-$\st$ coincide~\cite{dvorakcomplexity}. For this reason, the tests for both of these problems are shown in Figure~\ref{fig:dccob2dcprb2}\subref{fig:dccob2} by using \BA ($100$ ``YES'') and $\BC$ ($100$ ``NO''. Even in this case, in the case of ``NO'' QArg reaches the timeout because no solution can be (correctly) found.

Still, according to \cite{dvorakcomplexity}, for $\pr$ and $\sst$ semantics, the $\nem$ task reduces to check whether there is a non-empty admissible set (or complete extension). Thus, the tests of  $\ex$-$\co$, $\ex$-$\pr$, $\ex$-$\sst$, $\nem$-$\co$, $\nem$-$\pr$, and $\nem$-$\sst$  are collectively reported in Figure~\ref{fig:dccob2dcprb2}\subref{fig:dcprb2}. On \BA, we provide a comparison between checking $\nem$-$\ad$ and $\nem$-$\co$ (which leads to more timeouts than $\nem$-$\ad$ in general), while on \BD, we tested $\nem$-$\ad$ only.

In conclusion, the experiments in this section can be summarized in a few words. As approximate solvers, both AFGCN and Harper++ return many incorrect answers (particularly in the case of ``NO''). At the same time, QArg is correct but slower than the other solvers. However, as discussed in Section~\ref{sec:conclusion}, its speed can certainly be improved.

\begin{table}[t]
\begin{tabular}{cccccc}
\#Frameworks   & \#Tests        & Correct tests  & Tmax Exec (s) & Tmin Exec (s) & Tmed Exec (s) \\ 
100            & 500            & 474/500        & 0.634         & 0.15          & 0.296         \\ \hline
Tmax Comp (s)  & Tmin Comp (s)  & Tmed Comp (s)  & Max setcard    & Min setcard    & Med setcard    \\
142.579        & 2.574          & 36.316         & 80            & 1             & 40.5          \\ \hline
Max add        & Min add        & Med add        & Max remov     & Min remov     & Med remov     \\
2953           & 0              & 556            & 407           & 81            & 184.5         \\ \hline
Max Totaldiffs & Min Totaldiffs & Med Totaldiffs & Max atts      & Min atts      & Median atts   \\
3104           & 297            & 730            & 436           & 271           & 349.5        
\end{tabular}
\caption{Results on set \BB. The cells respectively report the number of tested AFs in \BB, the overall number of tests (frameworks times the number of tested argument sets on each framework, i.e., $5$), the number of solutions that are complete, the maximum, and minimum, median time taken to execute the QUBO model and to compile the model, the maximum, minimum, median cardinality of the sets randomly selected to be enforced, the maximum, minimum, median number of additions/removal/addition+removals concerning the original framework after the enforcement, and finally a description about the number of attacks in the dataset.}\label{tab:resenforcement1}
\end{table}

\subsection{Tests on the Extension Enforcement Problem}\label{sect:testextenforc}
We describe here the tests we performed to evaluate the QUBO encodings presented in Section~\ref{sect:enforcenc}. As mentioned in the introduction of Section~\ref{sec:implementation}, in this case, we are not aware of any public solver to be used for comparison; in addition, no specific benchmark has been assembled for this kind of enforcement problem. Moreover, as in Section~\ref{sec:comparison}, we use \emph{ConArg}~\cite{conarg,conarg2} as the reference solver to check if a set of arguments belongs to a given semantics or not.

Our tests consider the \BB dataset we introduced in the previous section. It consists of $100$ frameworks with $80$ arguments each, generated as \emph{Erd\H{o}s-R\'{e}nyi} (\emph{ER}) graphs~\cite{erdos}. For each AF of this benchmark, we generated five different (uniformly distributed) random sets of arguments to be (strictly) enforced as complete extensions; we discarded random sets that were already complete. The size of each set was randomly selected using a uniform distribution as well, between $1$ and the cardinality of the set of arguments in that AF. We set the penalty amplification value $\lambda$ to $100$ for breaking the completeness constraints (see the end of Section~\ref{sect:enforcenc}). In this case, no timeout was imposed; a summary of these tests is reported in Table~\ref{tab:resenforcement1}, where the caption describes each indicator. The number of correct tests is $476$ out of $500$: in $26$ cases, the modification to the framework did not force the requested set of arguments to be complete.

\section{Tests on a Quantum Machine}\label{sec:implementationQ}
In this section, we focus on testing the solution of our QUBO encoding on a quantum annealer offered by D-Wave Systems. This machine corresponds to the \emph{Advantage\_System 4.1 Quantum Annealer} (in the following of this paper, we abbreviate it to \emph{QA}), which employs a \emph{Pegasus} architecture/topology that has several improvements over its predecessors, i.e., \emph{Chimera}: for example, it allows more efficient embedding of cliques, bicliques, 3D lattices, and penalty models, and improved heuristic embedding run times~\cite{techdwave}. Moreover, it provides parity/auxiliary qubits, which are useful in encoding various logical constraints. This solver has $5760$ qubits and $15$ interactions per qubit~\cite{techdwave}. 

To encode the problems, we used the same D-Wave Ocean SDK adopted in Section~\ref{sec:implementation} to import the SA algorithm. In this case, the SDK libraries connect to the QA and send each problem to the online solver. The Ocean SDK includes a suite of open source Python tools\footnote{D-Wave Ocean SDK: \url{https://github.com/dwavesystems/dwave-ocean-sdk}.} for solving hard problems with local solvers, such as SA algorithms (as in Section~\ref{sec:implementation}), but also D-Wave's solvers by using the Leap Quantum Cloud Service\footnote{Leap Cloud Quantum: \url{https://cloud.dwavesys.com/leap/login/?next=/leap/}.}. 
To create the QA benchmark, we generated $20$ new AFs composed of $5$ AFs for each value of the number $n$ of arguments in the set $\{10, 15, 20, 25\}$. The number of arguments is smaller than the SA experiments in Section~\ref{sec:implementation} because we expected (as is actually) that QA would encounter difficulties even in solving problems of this size. These AFs were generated as \emph{Erd\H{o}s-R\'{e}nyi} graphs~\cite{erdos}, following the same methodology as in Section~\ref{sec:implementation}. Table~\ref{tab:instances} provides some statistics for each generated graph.

\begin{table}
\centering
\scalebox{0.83}{
\begin{tabular}{p{1cm}p{0.28cm}p{0.3cm}p{0.28cm}p{0.28cm}p{0.28cm}p{0.28cm}p{0.28cm}p{0.28cm}p{0.28cm}p{0.28cm}p{0.28cm}p{0.28cm}p{0.28cm}p{0.28cm}p{0.28cm}p{0.28cm}p{0.28cm}p{0.28cm}p{0.28cm}p{0.28cm}p{0.28cm}}
& \rotatebox{45}{af10\_1} & \rotatebox{45}{af10\_2} & \rotatebox{45}{af10\_3} & \rotatebox{45}{af10\_4} & \rotatebox{45}{af10\_5} & \rotatebox{45}{af15\_1} & \rotatebox{45}{af15\_2} & \rotatebox{45}{af15\_3} & \rotatebox{45}{af15\_4} & \rotatebox{45}{af15\_5} & \rotatebox{45}{af20\_1} & \rotatebox{45}{af20\_2} & \rotatebox{45}{af20\_3} & \rotatebox{45}{af20\_4} & \rotatebox{45}{af20\_5} & \rotatebox{45}{af25\_1} & \rotatebox{45}{af25\_2} & \rotatebox{45}{af25\_3} & \rotatebox{45}{af25\_4} & \rotatebox{45}{af25\_5}\\
\midrule
\#Args & 10 & 10 & 10 & 10 & 10 & 15 & 15 & 15 & 15 & 15 & 20 & 20 & 20 & 20 & 20 & 25 & 25 & 25 & 25 & 25 \\
\#Atts.& 13 & 13 & 20 & 14 & 19 & 21 & 29 & 30 & 27 & 28 & 51 & 52 & 51 & 53 & 45 & 78 & 76 & 80 & 81 & 73 \\
\#Complete & 2 & 3 & 4 & 6 & 3 & 3 & 3 & 3 & 3 & 3 & 3 & 6 & 2 & 5 & 3 & 7 & 3 & 6 & 3 & 3\\
\#Vars. & 20  &  24  &  32  &  20  &  30  &  35  &  47  &  51  &  45  &  45  &  84  &  88  &  84  &  88  &  74  &  131  &  129  &  137  &  139  &  123 \\
\end{tabular}
}
\caption{The AF instances generated to test the QA solver. Each instance is named in the table's header, and it has the number of arguments, the number of attacks, the number of complete extensions in each AF, and the number of binary variables in the QUBO encoding as defined in the four rows in the table.}\label{tab:instances}
\end{table}

In Section~\ref{sec:quantumtestscred}, we use these instances to test the credulous acceptance of an argument in complete extensions and the results obtained on enforcement problems.

\subsection{Results}\label{sec:quantumtestscred}
We tested the credulous acceptance of three different arguments for each AF in the QA benchmark.
These arguments were selected in the following way: \emph{i)} one argument that appears in only one of the complete extensions of the considered AF, \emph{ii)} one argument that appears in all the complete extensions (i.e., the argument is skeptically accepted, see Section~\ref{sec:background}), and \emph{iii)} one argument that is never accepted in complete extensions. Finding a solution for case \emph{i} should be more complicated than case \emph{ii}, while in both cases, the minimum energy value of the solution will be $0$.
In case \emph{iii}, the energy value of the best solution found must be greater than $0$.

For each couple $\langle$AF, argument$\rangle$ tested, we executed $10$ runs if the framework had $10$, $15$, $20$ arguments, and $20$ runs in the case of AFs with $25$ nodes because preliminary tests showed more variability with the latter class of AFs. In each run, the QA performed $1000$ reads.

Table~\ref{tab:testone} reports the results for case \emph{i}, that is, when the selected argument appears only in one extension. For each AF, corresponding to a different row in the table, we show the average energy obtained on reads with minimal energy (\emph{min\_energy}), the average energy of all the reads (\emph{avg\_energy}), the average number of reads in which the energy was equal to the minimum energy achieved (\emph{min\_energy\_freq}), the ratio between the number of qubits used by the annealer and the number of binary variables of the quadratic model (\emph{embedd\_overhead}), and the fraction of broken chains (\emph{chain\_break\_fract}).\footnote{When using a D-Wave QA, problems are mapped onto a physical hardware graph which may have a different topology from the logical problem graph. To represent logical variables that are connected in ways that are not supported directly by the hardware, a process called \emph{embedding} is used. This embedding involves chaining together multiple physical qubits to represent a single logical qubit. The fraction of chain breaks indicates how often these chains fail to stay connected during the annealing process (that is, the physical qubits that correspond to the same logical qubit assume different values), which can affect the quality of the solution~\cite{techdwave}.}
We also report the standard deviation of the same metrics on the right of each mean value. For each different set of AFs with the same number of arguments, we also aggregate the results together as an additional row of the table. For example, the table row \emph{all af10} aggregates the results for all AFs with $10$ arguments. The column labeled \emph{solved} shows the percentage of successful runs, i.e., the runs where at least one read achieved zero energy.

As expected, as the number of arguments $n$ increases, the results obtained by the QA are worse. Indeed, the results are excellent with $n=10$ arguments, as nearly all the reads in each run achieve an energy of $0$. The results when $n=15$ are slightly worse than those for $n=10$. A notable decrease in performance is observed when $n=20$. It is possible to notice that the behavior of QA changes dramatically when $n=25$. Specifically, two distinct phenomena must be addressed: first, the minimum energy found in each run is not always $0$; in fact, the percentage of successful runs is less than $100\%$ across all AFs. Second, the number of reads for which the minimum energy is found is very small: about $3$-$5$ reads out of $1000$.

As expected, the overhead produced by the embedder constantly increases with the number of arguments $n$. For $n=25$ the QA uses a number of qubits that is almost three times the number of variables in the corresponding QUBO problem. Moreover, the fraction of broken chains also increases.

\begin{table}
\centering
\scalebox{0.85}{
\begin{tabular}{llr|lr|lr|c|lr|lr}
\toprule
 & \multicolumn{2}{c}{\textbf{min\_energy}} & \multicolumn{2}{c}{\textbf{avg\_energy}} & \multicolumn{2}{c}{\textbf{min\_energy\_freq}} & \textbf{solved} & \multicolumn{2}{c}{\textbf{embedd\_overhead}} & \multicolumn{2}{c}{\textbf{chain\_break\_fract}}\\
 & mean & std & mean & std & mean & std & \% & mean & std & mean & std \\
AF &  &  &  &  &  &  &  &  &  &  &  \\
\midrule
af10\_1 & 0.000 & 0.000 & 0.033 & 0.017 & 974.600 & 16.507 & 100 & 1.221 & 0.054 & 0.011 & 0.022\\
af10\_2 & 0.000 & 0.000 &  0.050 & 0.038 & 953.700 & 37.044 & 100 & 1.183 & 0.040 & 0.004 & 0.014\\
af10\_3 & 0.000 & 0.000 & 0.337 & 0.106 & 715.600 & 87.428 & 100 & 1.368 & 0.091 & 0.023 & 0.034\\
af10\_4 & 0.000 & 0.000 & 0.123 & 0.057 & 890.400 & 52.282 & 100 & 1.263 & 0.086 & 0.021 & 0.037\\
af10\_5 & 0.000 & 0.000 & 0.193 & 0.107 & 832.900 & 87.872 & 100 & 1.407 & 0.051 & 0.017 & 0.024\\
\hline
all af10 & 0.000 & 0.000 & 0.147 & 0.133 & 873.440 & 111.838 & 100 & 1.288 & 0.108 & 0.015 & 0.027\\
\hline
\hline
af15\_1 & 0.000 & 0.000 &  0.123 & 0.053 & 901.000 & 44.337 & 100 & 1.432 & 0.080 & 0.032 & 0.026\\
af15\_2 & 0.000 & 0.000 & 0.292 & 0.147 & 784.100 & 111.726 & 100 & 1.596 & 0.093 & 0.043 & 0.043\\
af15\_3 & 0.000 & 0.000 & 0.713 & 0.445 & 629.500 & 148.278 & 100 & 1.730 & 0.102 & 0.376 & 0.918\\
af15\_4 & 0.000 & 0.000 & 0.065 & 0.035 & 956.400 & 20.866 & 100 & 1.400 & 0.090 & 0.020 & 0.023\\
af15\_5 & 0.000 & 0.000 & 0.595 & 0.228  & 548.200 & 164.442 & 100 & 1.482 & 0.052 & 0.064 & 0.053\\
\hline
all af15 & 0.000 & 0.000 & 0.357 & 0.343  & 763.840 & 190.794 & 100 & 1.528 & 0.147 & 0.107 & 0.418\\
\hline
\hline
af20\_1 & 0.000 & 0.000 & 2.272 & 0.824 & 156.700 & 104.691 & 100 & 2.233 & 0.192 & 0.505 & 0.481\\
af20\_2 & 0.000 & 0.000 & 2.388 & 0.620 & 167.800 & 112.956 & 100 & 2.271 & 0.176 & 0.352 & 0.268\\
af20\_3 & 0.000 & 0.000 & 2.617 & 0.638 & 79.000 & 66.148 & 100 & 2.283 & 0.093 & 0.782 & 0.497\\
af20\_4 & 0.000 & 0.000 & 2.623 & 0.673 & 169.700 & 75.504 & 100 & 2.301 & 0.143 & 0.922 & 0.887\\
af20\_5 & 0.000 & 0.000 & 1.481 & 0.637 & 298.200 & 173.536 & 100 & 1.960 & 0.140 & 0.521 & 0.428\\
\hline
all af20 & 0.000 & 0.000 &  2.276 & 0.780 & 174.280 & 129.619 & 100 & 2.210 & 0.194 & 0.616 & 0.568\\
\hline
\hline
af25\_1 & 0.300 & 0.470 & 7.057 & 0.682 & 3.750 & 2.845 & 70 & 3.013 & 0.227 & 2.105 & 1.248\\
af25\_2 & 0.350 & 0.587 & 7.174 & 1.045  & 5.600 & 5.452 & 70 & 2.753 & 0.138 & 2.696 & 2.288\\
af25\_3 & 0.900 & 0.968 & 7.682 & 0.932 & 3.950 & 3.980 & 45 & 2.829 & 0.194 & 3.082 & 1.611\\
af25\_4 & 0.600 & 0.598 & 7.596 & 1.127 & 3.850 & 4.749 & 45 & 3.004 & 0.176 & 3.795 & 2.012\\
af25\_5 & 0.650 & 0.745 & 6.190 & 0.961  & 4.150 & 4.107 & 50 & 2.758 & 0.167 & 1.791 & 1.264\\
\hline
all af25 & 0.560 & 0.715 &  7.140 & 1.083  & 4.260 & 4.282 & 56 & 2.871 & 0.213 & 2.694 & 1.843\\
\hline
\hline
\end{tabular}
}
\caption{Results obtained when the selected argument appears in one complete extension only.}\label{tab:testone}
\end{table}

Table~\ref{tab:testmany} reports the results for case \emph{ii}, i.e., when we select an argument that appears in all the complete extensions. Consequently, the search phase can reach more solutions. The findings slightly outperform those in Table~\ref{tab:testone}, and similar observations can be made.

\begin{table}
\centering
\scalebox{0.85}{
\begin{tabular}{llr|lr|lr|c|lr|lr}
\toprule
 & \multicolumn{2}{c}{\textbf{min\_energy}} & \multicolumn{2}{c}{\textbf{avg\_energy}} & \multicolumn{2}{c}{\textbf{min\_erg\_freq}} & \textbf{solved} & \multicolumn{2}{c}{\textbf{embedd\_overhead}} & \multicolumn{2}{c}{\textbf{chain\_break\_fract}}\\
 & mean & std & mean & std & mean & std & \% & mean & std & mean & std \\
AF &  &  &  &  &  &  &  &  &  &  &  \\
\midrule
af10\_1 & 0.000 & 0.000 & 0.050 & 0.022 & 960.300 & 19.556 & 100 & 1.332 & 0.086 & 0.016 & 0.025 \\
af10\_2 & 0.000 & 0.000 & 0.060 & 0.029 & 944.800 & 25.862 & 100 & 1.274 & 0.054 & 0.004 & 0.014 \\
af10\_3 & 0.000 & 0.000 & 0.216 & 0.070 & 814.400 & 58.155 & 100 & 1.526 & 0.083 & 0.126 & 0.297 \\
af10\_4 & 0.000 & 0.000 & 0.031 & 0.011 & 972.500 & 10.233 & 100 & 1.232 & 0.051 & 0.011 & 0.022 \\
af10\_5 & 0.000 & 0.000 & 0.223 & 0.161 & 820.700 & 118.139 & 100 & 1.390 & 0.069 & 0.076 & 0.109 \\
\hline
all af10 & 0.000 & 0.000 & 0.116 & 0.115 & 902.540 & 91.616 & 100 & 1.351 & 0.124 & 0.046 & 0.144 \\
\hline
\hline
af15\_1 & 0.000 & 0.000 & 0.134 & 0.037 & 892.400 & 29.500 & 100 & 1.506 & 0.074 & 0.035 & 0.036 \\
af15\_2 & 0.000 & 0.000 & 0.186 & 0.083 & 862.600 & 62.491 & 100 & 1.535 & 0.046 & 0.046 & 0.061 \\
af15\_3 & 0.000 & 0.000 & 0.323 & 0.098 & 799.100 & 47.302 & 100 & 1.656 & 0.085 & 0.204 & 0.319 \\
af15\_4 & 0.000 & 0.000 & 0.212 & 0.088 & 843.100 & 61.557 & 100 & 1.536 & 0.088 & 0.120 & 0.102 \\
af15\_5 & 0.000 & 0.000 & 0.825 & 0.210 & 393.900 & 112.313 & 100 & 1.673 & 0.140 & 0.370 & 0.934 \\
\hline
all af15 & 0.000 & 0.000 & 0.336 & 0.27 & 758.220 & 197.732 & 100 & 1.581 & 0.112 & 0.155 & 0.444 \\
\hline
\hline
af20\_1 & 0.000 & 0.000 & 1.868 & 0.440 & 178.100 & 83.945 & 100 & 2.207 & 0.137 & 0.331 & 0.135 \\
af20\_2 & 0.000 & 0.000 & 2.202 & 0.529 & 157.000 & 98.071 & 100 & 2.343 & 0.176 & 1.257 & 1.614 \\
af20\_3 & 0.000 & 0.000 & 2.460 & 0.346 & 127.500 & 97.716 & 100 & 2.298 & 0.060 & 1.684 & 1.612 \\
af20\_4 & 0.000 & 0.000 & 2.534 & 0.361 & 78.200 & 34.985 & 100 & 2.369 & 0.146 & 1.811 & 1.903 \\
af20\_5 & 0.000 & 0.000 & 0.793 & 0.154 & 489.900 & 66.734 & 100 & 1.911 & 0.088 & 0.156 & 0.183 \\
\hline
all af20 & 0.000 & 0.000 & 1.972 & 0.739 & 206.140 & 165.970 & 100 & 2.225 & 0.209 & 1.048 & 1.452 \\
\hline
\hline
af25\_1 & 0.100 & 0.447 & 6.292 & 0.886 & 6.300 & 4.330 & 95 & 3.080 & 0.155 & 1.898 & 1.319 \\
af25\_2 & 0.850 & 0.587 & 7.007 & 0.609 & 3.300 & 2.774 & 25 & 2.850 & 0.207 & 2.562 & 1.278 \\
af25\_3 & 0.350 & 0.671 & 6.571 & 0.750 & 5.650 & 4.234 & 75 & 2.799 & 0.115 & 1.977 & 1.005 \\
af25\_4 & 0.600 & 0.503 & 6.858 & 0.747  & 4.250 & 3.177 & 40 & 2.968 & 0.229 & 3.459 & 1.802 \\
af25\_5 & 0.550 & 0.686 & 6.536 & 1.092 & 4.950 & 4.123 & 55 & 2.732 & 0.152 & 2.013 & 1.233 \\
\hline
all af25 & 0.490 & 0.628 & 6.653 & 0.854  & 4.890 & 3.850 & 58 & 2.886 & 0.213 & 2.382 & 1.451 \\
\hline
\hline
\end{tabular}
}
\caption{Results obtained when the selected argument appears in all the complete extensions.}\label{tab:testmany}
\end{table}

Then, Table~\ref{tab:testnone} shows the results for case \emph{iii}. Under these circumstances, the minimum energy correctly never reaches $0$ as the tested problem has no solution. This is evident from the average of the minimum energy values (\emph{min\_energy}). For the same reason, Table~\ref{tab:testnone} has no column \emph{solved}.
Nevertheless, we can still count the number of reads that reached the lowest possible energy value for this problem. Similarly to what happens for the cases \emph{i} and \emph{ii}, this number is very large for $n=10$, slightly decreases for $n=15$ and sharply for $n=20$, while for $n=25$ it is just around $3$-$5$ reads per run.

\begin{table}
\centering
\scalebox{0.9}{
\begin{tabular}{llr|lr|lr|lr|lr}
\hline
 & \multicolumn{2}{c}{\textbf{min\_energy}} & \multicolumn{2}{c}{\textbf{avg\_energy}} & \multicolumn{2}{c}{\textbf{freq\_min\_erg}} & \multicolumn{2}{c}{\textbf{embedd\_overhead}} & \multicolumn{2}{c}{\textbf{chain\_break\_fract.}} \\
 & mean & std & mean & std & mean & std & mean & std & mean & std \\
AF &  &  &  &  &  &  &  &  \\
\midrule
af10\_1 & 2.000 & 0.000 & 2.020 & 0.008 & 981.400 & 8.195 & 1.247 & 0.050 & 0.021 & 0.027 \\
af10\_2 & 2.000 & 0.000 & 2.036 & 0.015 & 966.600 & 14.864 & 1.222 & 0.038 & 0.009 & 0.018 \\
af10\_3 & 2.000 & 0.000 & 2.294 & 0.153 & 757.300 & 121.101 & 1.481 & 0.115 & 0.210 & 0.369 \\
af10\_4 & 2.000 & 0.000 & 2.027 & 0.008 & 976.300 & 7.484 & 1.211 & 0.050 & 0.021 & 0.027 \\
af10\_5 & 2.000 & 0.000 & 2.274 & 0.126 & 750.600 & 115.133 & 1.414 & 0.054 & 0.097 & 0.143 \\
\hline
all af10 & 2.000 & 0.000 &  2.130 & 0.153 & 886.440 & 130.998 & 1.315 & 0.129 & 0.071 & 0.187 \\
\hline
\hline
af15\_1 & 2.000 & 0.000 & 2.402 & 0.231 & 679.700 & 175.547 & 1.471 & 0.095 & 0.065 & 0.091 \\
af15\_2 & 3.000 & 0.000 & 3.676 & 0.163 & 499.000 & 103.542 & 1.554 & 0.083 & 0.200 & 0.280 \\
af15\_3 & 2.000 & 0.000 & 2.580 & 0.305 & 661.100 & 132.219 & 1.772 & 0.078 & 0.224 & 0.181 \\
af15\_4 & 2.000 & 0.000 & 2.432 & 0.134 & 677.400 & 94.312 & 1.518 & 0.077 & 0.023 & 0.021 \\
af15\_5 & 4.000 & 0.000 & 4.416 & 0.181 & 676.300 & 122.878 & 1.477 & 0.065 & 0.309 & 0.350 \\
\hline
all af15 & 2.600 & 0.808 & 3.101 & 0.841 & 638.700 & 142.372 & 1.558 & 0.136 & 0.164 & 0.237 \\
\hline
\hline
af20\_1 & 2.000 & 0.000 & 4.177 & 0.361 & 103.700 & 51.203 & 2.036 & 0.083 & 0.912 & 0.827 \\
af20\_2 & 3.000 & 0.000 & 6.180 & 0.609 & 43.600 & 35.078 & 2.270 & 0.125 & 2.734 & 2.448 \\
af20\_3 & 2.000 & 0.000 &  4.578 & 0.346 & 90.100 & 40.23 & 2.329 & 0.139 & 1.602 & 2.045 \\
af20\_4 & 2.000 & 0.000 & 4.495 & 0.432 & 63.600 & 48.427 & 2.180 & 0.124 & 1.218 & 1.107 \\
af20\_5 & 2.000 & 0.000 & 3.620 & 0.626 & 277.500 & 130.593 & 2.095 & 0.144 & 0.929 & 1.055 \\
\hline
all af20 & 2.200 & 0.404 & 4.610 & 0.982 & 115.700 & 108.115 & 2.182 & 0.162 & 1.479 & 1.700 \\
\hline
\hline
af25\_1 & 1.500 & 0.607 & 7.677 & 1.019 & 3.900 & 3.972 & 3.050 & 0.196 & 2.343 & 1.813 \\
af25\_2 & 1.400 & 0.503 & 7.244 & 0.689 & 5.100 & 6.138 & 2.766 & 0.159 & 2.652 & 1.729 \\
af25\_3 & 2.200 & 0.410 & 8.293 & 0.725 & 6.000 & 3.798 & 2.840 & 0.187 & 2.218 & 1.167 \\
af25\_4 & 2.100 & 0.308 & 8.337 & 1.158 & 3.800 & 4.663 & 2.913 & 0.154 & 4.347 & 3.080 \\
af25\_5 & 1.250 & 0.444 & 6.790 & 0.850 & 4.600 & 3.619 & 2.702 & 0.146 & 1.683 & 1.701 \\
\hline
all af25 & 1.690 & 0.598 & 7.669 & 1.072 & 4.680 & 4.515 & 2.854 & 0.206 & 2.648 & 2.161 \\
\hline
\hline
\end{tabular}
}
\caption{Results obtained when the selected argument is never accepted in a complete extension.}\label{tab:testnone}
\end{table}

Finally, Table~\ref{tab:enforcementquantum} shows the results we obtained with the (strict) enforcement of complete extensions.  In this round of tests, for each $10$-node AF previously introduced, we generated a set of arguments that need only one modification to be enforced as a complete extension (i.e., one attack to be added/removed to the AF).

We only used AFs with arguments $10$ because of the difficulty we expected due to the increased number of variables to model the corresponding QUBO problem. We have computed that encoding the enforcement problem as QUBO requires around $250$ variables, which is
almost twice the number of binary variables required by the largest AF for the credulous acceptance problem, as shown in Table~\ref{tab:instances}.

This expectation is confirmed by more than one column in the table: for example, on \emph{af10\_2}, finding a solution in two rounds of $1000$ reads was impossible. Moreover, the chain break fraction is markedly higher than the credulous acceptance problems: about twice the same value for $25$-node AFs and the credulous acceptance problem in previous tables. However, the embedding overhead is not that significantly high.

The average value for minimal energy solutions (column \emph{min\_energy}) correctly reports the penalty of $1$ to pay to add/remove one attack, clearly except for \emph{af10\_2}.
The column \emph{sols\_num} indicates the min/mean/max number of reads where a solution of energy $1$ corresponds to the correct enforcement.

\begin{table}
\centering
\scalebox{0.83}{
\begin{tabular}{llr|lr|lrr|c|lr|lr}
\toprule
  & \multicolumn{2}{c}{\textbf{min\_energy}} & \multicolumn{2}{c}{\textbf{avg\_energy}} & \multicolumn{3}{c}{\textbf{sols\_num}} & \textbf{solved} & \multicolumn{2}{c}{\textbf{embedd\_overhead}} & \multicolumn{2}{c}{\textbf{chain\_break\_fraction}} \\
 & mean & std & mean & std & min & mean & max & sum & mean & std & mean & std \\
AF &  &  &  &  &  &  &  & \% &  &  &  &  \\
\midrule
af10\_1 & 1.000 & 0.000 & 4.345 & 1.095 & 1 & 59.400 & 173 & 100 & 1.620 & 0.087 & 6.105 & 4.033 \\
af10\_2 & 1.300 & 0.675 & 5.653 & 0.992 & 0 & 7.800 & 44 & 80 & 1.729 & 0.101 & 7.397 & 2.392 \\
af10\_3 & 1.000 & 0.000 & 4.969 & 1.405 & 1 & 30.300 & 129 & 100 & 1.729 & 0.069 & 5.240 & 2.147 \\
af10\_4 & 1.000 & 0.000 & 4.889 & 1.224 & 1 & 44.100 & 146 & 100 & 1.784 & 0.100 & 4.947 & 3.390 \\
af10\_5 & 1.000 & 0.000 & 5.258 & 0.865 & 4 & 31.600 & 73 & 100 & 1.738 & 0.108 & 5.572 & 1.535 \\
\hline
\hline
\end{tabular}
}
\caption{Results obtained on the strict enforcement of a set of arguments as a complete extension.}\label{tab:enforcementquantum}
\end{table}

\section{Related Work}\label{sec:related}

In the literature, many computational techniques and practical implementations are available to solve problems related to formal argumentation. This section will focus on two methods: the \emph{reduction-based} and \emph{direct} approaches. Reduction-based approaches involve reducing the problem into a different formalism to take advantage of existing solvers from that formalization. On the other hand, direct approaches involve designing specific algorithms to solve the problem directly.

The \emph{International Competition on Computational Models of Argumentation} (\emph{ICCMA} for short)\footnote{ICCMA Website: \url{http://argumentationcompetition.org}.} is the reference biennial competition dedicated to argumentation, whose objectives are to provide a forum for the empirical comparison of solvers, to highlight challenges to the community, proposing new directions for research, and providing a core of common benchmark instances and a representation formalism that can aid in the comparison and solver evaluation.

In the following, we report some of the complete solvers that participated in recent ICCMA competitions for the sake of completeness. At the same time, Section~\ref{sect:approximatesolvers} will focus on the works most related to what this paper proposes, i.e., approximate solvers. 

For example, $\mu$-toksia~\cite{DBLP:conf/kr/NiskanenJ20a} is a purely SAT-based implementation: all the reasoning by the system is performed by calls to a \emph{Boolean satisfiability} (\emph{SAT}) solver, including polynomial-time computations such as the grounded semantics. Even \emph{CoQuiAAS}~\cite{DBLP:conf/ictai/LagniezLM15} adopts an encoding to SAT and then a search of \emph{Maximal Satisfiable Subsets} of constraints in a \emph{Partial Max-SAT} instance. 

The \emph{Pyglaf} reasoner~\cite{DBLP:journals/fuin/Alviano19} reduces the considered problem to \emph{circumscription} employing linear encodings.  Circumscription is a non-monotonic logic formalizing common sense reasoning employing a second-order semantics, which essentially enforces minimizing the extension of some predicates. The circumscription solver extends the used SAT solver \emph{glucose} and implements an algorithm based on unsatisfiable core analysis.

\emph{Argpref}~\cite{DBLP:conf/sac/PrevitiJ18} is a solver specialized in computing the ideal semantics. It implements a SAT-with-preferences approach to computing the backbone of a propositional encoding of admissible sets. Then, polynomial-time post-processing is applied to construct the ideal extension.

\emph{Aspartix-V21}~\cite{DvorakKWW21} is an evolution of one of the first solvers for AFs, i.e., \cite{eglyetal2008bis}. It is based on an encoding of problems by using \emph{Answer Set Programming} (\emph{ASP}), and it has been extended to deal with several formalisms and extensions of AFs (e.g., preferences associated with arguments).

ConArg~\cite{conarg2} encodes problems as \emph{Constraint Satisfaction Problems} (\emph{CSPs}), and it has been used as the reference solver of ICCMA'19. Successively, the features of this solver have been proposed as a C++ library~\cite{conarg} providing an API to solve Argumentation problems such as, for example, enumeration, existence, acceptance, verification, and non-emptyness.

Although all the works presented so far follow a reduction-based approach in SAT, ASP, CSP encodings,  \cite{DBLP:journals/ijar/NofalAD16} is an example of a direct approach: the authors enhance backtracking search for sets of acceptable arguments by using a more powerful pruning strategy, named the \emph{global looking-ahead strategy}.

\subsection{Approximate Solvers}\label{sect:approximatesolvers}
We point the interested reader to the survey of participants and the results achieved in ICCMA15~\cite{iccma15}, ICCMA17~\cite{iccma17}, and ICCMA19~\cite{iccma19}. 
The last edition of the competition, ICCMA21, saw the participation of nine solvers, and it confirmed a third class of solvers in addition to reduction-based and direct techniques: approximate approaches. An exploratory track dedicated to such algorithms was included for the first time: only decision problems $\dc$-$\sigma$ and $\ds$-$\sigma$ were considered for five different sub-tracks, such as $\sigma \in \{ \co, \pr, \st, \sst, \stg \}$, as well as $\ds$-$\id$. Solvers were evaluated concerning accuracy, i.e., correctly solved instances' ratios. The main motivation behind approximate algorithms over exact algorithms was their (potentially) lower execution: the timeout was reduced to 60 seconds instead of 600.

An approximate solver from ICCMA21 is HARPER++ by M. Thimm: such a solver can only determine the grounded extension of an input framework and then uses that to approximate the results for $\dc$ and $\ds$ tasks related to $\sigma \in \{\co, \st, \pr, \sst, \stg, \id\}$. A positive answer to $\ds$-$\gr$ implies a positive answer to $\dc$ and $\ds$ for the other semantics. In contrast, if an argument is attacked by an argument in the grounded extension, the answer to $\dc$ and $\ds$ is negative. According to \cite{CeruttiTV20}, skeptical reasoning with any semantics generally overlaps with reasoning with grounded semantics in many practical cases.

AFGCN, by Lars Malmqvist, also competed in ICCMA21. It uses a Graph Convolutional Network~\cite{WuPCLZY21}, to compute approximate solutions to $\dc$ and $\ds$ tasks for $\sigma \in \{\co, \st, \pr, \sst, \stg, \id\}$ in a given AF. The model is trained using a randomized training process using a dataset of AFs from previous ICCMA competitions to maximize the generalization of the input framework. To speed up calculation and improve accuracy, the solver uses the pre-computed grounded extension as an input feature to the neural network.

\section{Conclusions and Future Work}\label{sec:conclusion}

This paper provided a QUBO encoding of some NP-Complete problems in Abstract Argumentation Frameworks for the first time. From all the tests we performed locally using the SA algorithm, the encoding is shown to be correct. The obtained results show better accuracy than the related approximate solvers in the literature. Moreover, we also performed an experimentation phase on a Quantum Annealer offered by D-Wave, sending small AFs to the solver and receiving a response for their solutions. The considered QA cannot already solve all the proposed framework instances with $25$ arguments (credulous acceptance in complete extensions problem) and  $10$ arguments (strict enforcement of complete extensions).  For larger instances, e.g., the $80$-argument AFs we used to test SA on a local machine, the D-Wave embedder in the Python library replied with the impossibility of mapping the problem to the Quantum Annealer.

To the best of our knowledge, the optimization behind mapping QUBO models derived from an Argumentation problem to the architecture of quantum machines is still unexplored and challenging: several parameters need further investigation to exploit better the hardware and the connections among qubits, which are limited by D-Wave's architecture. We need to leave this to future work. We additionally aim to perform more fine-tuning to optimize the search on the Quantum Annelaer just as we did locally with SA in Section~\ref{sec:implementation}. We need to acquire more computational time from D-Wave to do this, as the freely available resources are limited.


We also plan to use the \emph{Quantum Approximate Optimization Algorithm} (\emph{QAOA})  algorithm~\cite{qaoa} to solve argumentation tasks with gate-based quantum computers. The QUBO formulation can define the Hamiltonian problem and, hence, the parametric circuit for the QAOA approach.

Finally, in the future, we will extend the QUBO encoding to weighted problems in Argumentation~\cite{vbased} to represent weights (or probabilities) associated with arguments or attacks: this is allowed by using linear or quadratic coefficients that encode a weight in the expression modeling the problem. 

\section*{Acknowledgements}
 M. Baioletti was partially funded by University of
Perugia - "Fondo Ricerca di Ateneo WP4.4" 2022 - Project Quanta -
Laboratorio di Calcolo Quantistico.

F. Santini was partially funded by University of Perugia - "Fondo
Ricerca di Ateneo WP4.1" 2022 - Project RATIONALISTS, and by
European Union - Next Generation EU PNRR MUR PRIN - Project
J53D23007220006 EPICA: "Empowering Public Interest Communica-
tion with Argumentation".

\bibliography{main,biblio}

\end{document}